\documentclass[12pt,a4paper]{amsart}

\usepackage[latin9]{inputenc}
\usepackage[numbers, square, comma, compress]{natbib}
\usepackage{amsbsy}
\usepackage{amsfonts}
\usepackage{amsmath}
\usepackage{amssymb}
\usepackage{amstext}
\usepackage{amsthm}
\usepackage{appendix}
\usepackage{array}
\usepackage{bbm}
\usepackage{caption}
\usepackage{color}
\usepackage{eepic}
\usepackage{empheq}
\usepackage{epic}
\usepackage{epsfig}
\usepackage{eurosym}
\usepackage{float}
\usepackage{framed}
\usepackage{graphicx}
\usepackage{indentfirst}
\usepackage{latexsym}
\usepackage{longtable}
\usepackage{lscape}
\usepackage{marginnote}
\usepackage{pdfpages}
\usepackage{pgf}
\usepackage{pstricks}
\usepackage{pxfonts}
\usepackage{rotating}
\usepackage{subcaption}
\usepackage{tabularx}
\usepackage{tikz}
\usepackage{txfonts}
\usepackage{varioref}
\usepackage{wasysym}
\usepackage{textcomp}
\usepackage{yfonts}
\usepackage{soul}

\usepackage[scale=0.76]{geometry}

\makeatletter

\pdfpageheight\paperheight
\pdfpagewidth\paperwidth

\numberwithin{equation}{section}
\numberwithin{figure}{section}

\newcommand\ol[1]{{\setul{-0.9em}{}\ul{#1}}}

\numberwithin{equation}{section}
\numberwithin{figure}{section}

\def\ll{\left\lgroup}
\def\rr{\right\rgroup}

\def\leq{\leqslant}

\def\x{x^{(1)}}

\def\ll{ \left\lgroup}
\def\rr{\right\rgroup}

\newcommand{\cA}{\mathcal{A}}
\newcommand{\cB}{\mathcal{B}}
\newcommand{\cC}{\mathcal{C}}
\newcommand{\cD}{\mathcal{D}}

\newcommand{\cH}{\mathcal{H}}
\newcommand{\cI}{\mathcal{I}}

\newcommand{\cL}{\mathcal{L}}
\newcommand{\cM}{\mathcal{M}}

\newcommand{\cQ}{\mathcal{Q}}
\newcommand{\cR}{\mathcal{R}}
\newcommand{\cS}{\mathcal{S}}

\newcommand{\cU}{\mathcal{U}}

\newcommand{\CC}{\mathbbm{C}}

\newcommand{\NN}{\mathbbm{N}}

\newcommand{\RR}{\mathbbm{R}}
\newcommand{\ZZ}{\mathbbm{Z}}

\newcommand{\1}{{\bf 1.}}
\newcommand{\2}{{\bf 2.}}
\newcommand{\3}{{\bf 3.}}

\newcommand{\infinity}{\infty}

\newcommand{\bea}{\begin{eqnarray}\displaystyle}
\newcommand{\eea}{\end{eqnarray}}

\newcommand{\h}{{\bf h}}

\renewcommand{\H}{{\mathcal H}}

\def\ll{\left\lgroup}
\def\rr{\right\rgroup}

\restylefloat{figure}

\begin{document}

\title[Local height probabilities and critical partition functions]
{
Off-critical local height probabilities on a plane \\ 
and critical partition functions on a cylinder
} 

\author{Omar Foda}

\address{School of Mathematics and Statistics,
         University of Melbourne, 
         Parkville, Victoria 3010, Australia.}

\email{omar.foda@unimelb.edu.au}

\date{}

\begin{abstract}
We compute off-critical local height probabilities in regime-III restricted solid-on-solid models 
in a $4 N$-quadrant spiral geometry, with periodic boundary conditions in the angular direction, 
and fixed boundary conditions in the radial direction, as a function of $N$, the winding number 
of the spiral, and $\tau$, the departure from criticality of the model, and observe that the result 
depends only on the product $N \, \tau$.
In the limit $N \rightarrow 1$, $\tau \rightarrow \tau_0$, such that $\tau_0$ is finite, 
we recover the off-critical local height probability on a plane, $\tau_0$-away from criticality. 
In the limit $N \rightarrow \infinity$, $\tau \rightarrow 0$, such that $N \, \tau = \tau_0$ 
is finite, and following a conformal transformation, we obtain a critical partition function 
on a cylinder of aspect-ratio $\tau_0$. 
We conclude that the off-critical local height probability on a plane, $\tau_0$-away from 
criticality, is equal to a critical partition function on a cylinder of aspect-ratio $\tau_0$, 
in agreement with a result of Saleur and Bauer.
\end{abstract}

\maketitle

\section{Introduction}
\label{section.introduction}

\subsection{Background}
In 1987, Date, Jimbo, Miwa and Okado observed that (the essential part of) the \textit{off-critical} local 
height probabilities in an important class of solved statistical mechanical models on a square lattice are 
characters of fully-degenerate, irreducible Virasoro highest-weight modules, and 
in 1989, Saleur and Bauer observed that the same Virasoro characters are \textit{critical} partition 
functions of conformal field theories on a cylinder, if the departure from criticality in the off-critical 
local height probabilities of Date \textit{et al.} is matched to the aspect-ratio of the cylinder. 
In 1991, the present author proposed an argument to explain (or at least to motivate) the observation of 
Saleur and Bauer. 
This argument appeared in a preprint form, but was not widely-circulated and was not published. The present 
note recalls this argument. 

\subsection{Outline of contents} 
In section \textbf{\ref{section.restricted.solid.on.solid}}, we recall basic definitions related 
to the restricted solid-on-solid models of Andrews, Baxter and Forrester, and
in \textbf{\ref{section.corner.transfer.matrices.local.height.probabilities}}, we do the same for 
Baxter's corner transfer matrix method and local height probabilities.  
In \textbf{\ref{section.planar.geometry}}, we outline how the corner transfer matrix is used to 
compute off-critical local height probabilities on a plane, in the form of modular functions, 
and in \textbf{\ref{section.affine.virasoro.characters}}, we recall the re-writing of the off-critical 
local height probabilities in terms of affine and Virasoro characters
\footnote{\,
Sections \textbf{2--5} are descriptive guides to the original literature rather than a substitute for it. 
They contain just enough information, and notation, to allow the reader to follow the subsequent sections, 
and can be skipped by the expert reader.
}.
In \textbf{\ref{section.other.works}}, we recall the results of pioneering works where critical 
partition functions on a cylinder with fixed boundary conditions are expressed in terms of Virasoro 
characters, then outline the approach of the present work. 
In \textbf{\ref{section.spiral.geometry}}, we extend the regime-III restricted solid-on-solid local 
height probabilities to a $4N$-quadrant spiral geometry, with $N = 2, 3, \cdots$, and show that the 
result is the product of two factors, \textbf{A} and \textbf{B}, where 
\textbf{A} is a normalization that depends on $N$ and on $\tau$ separately, and 
\textbf{B} carries the statistical mechanics of the local height probability and depends only on 
the product $N \, \tau$.
We give up the normalization of the local height probabilities as probabilities for $N \neq 1$, 
rewrite them in terms of affine and Virasoro characters and formally set $N \in \RR$.
In \textbf{\ref{section.cylinder.geometry}}, we regularize the $4N$-quadrant off-critical local 
height probability, and take $N \rightarrow \infty$, $\tau \rightarrow 0$, such that $N \, \tau$ 
is fixed. In the limit, \textbf{A} simplifies, while \textbf{B} remains \textit{invariant}, and 
the system is critical on a spiral geometry with infinite winding number. Conformally transforming 
the spiral to a cylinder, we obtain a critical partition functions on a cylinder.
In section \textbf{\ref{section.postscript.comments}}, we collect a number of comments. 

\subsection{Abbreviations}
To simplify the presentation, we often use 
\1 
\lq local height probability\rq\, for 
\lq off-critical local height probability on a plane\rq, as one normally computes the local 
height probabilities off-criticality and on a plane, unless otherwise specified, 
\2
\lq Virasoro highest weight module\rq\, for 
\lq fully-degenerate irreducible Virasoro highest-weight module\rq, since we focus on statistical 
mechanical models that reduce at criticality to minimal conformal field theories based on Virasoro, 
rather than higher rank algebras.  

\subsection{Notation}
In a restricted solid-on-solid model $\cL_{\, m, \, m+1}$, $m = 3, 4, \cdots$, is the number 
of allowed heights. 
The parameter $\tau \in \RR$ measures the departure of an off-critical system from criticality, 
and $\tau^{\, \prime} = 1 / \tau$. 
The off-critical restricted solid-on-solid model $\cL_{\, m, \, m+1}$, $m = 3, 4, \cdots$, 
is defined in terms of Boltzmann weights parameterized by elliptic theta functions. The nome
of these theta functions is $p = e^{\, - \, 2 \, \pi \tau^{\, \prime}}$. 
The local height probabilities, which are essentially Virasoro characters, are $q$-series in 
$q = e^{\, - \, 4 \, \pi \, \tau / \ll m+1 \rr}$. 
When working on a spiral geometry of winding number $N$, we take the modular parameter $\tau$,
and consequently the nome $p$ and its modular conjugate $q$, 
to depend on $N$, and $\tau_{\, N}$, $p_{\, N}$ and $q_{\, N}$. 

\section{Restricted solid-on-solid models}
\label{section.restricted.solid.on.solid}
\textit{We recall basic facts related to the restricted solid-on-solid models of Andrews,
Baxter and Forrester.}

\subsection{The hard-square model} In \cite{baxter.1980, baxter.1981}, Baxter solved
the hard-square model on a square lattice, where nearest-neigh\-bour\-ing sites cannot 
be simultaneously occupied, and interactions are along the diagonals. 
The hard-hexagon model on a triangular lattice is recovered by taking the energy of the interactions 
along one of the two diagonals to be infinite, so that the two sites connected by that diagonal cannot 
be simultaneously occupied. If we think of the free/occupied sites as carrying a height variable 
$h = 0/1$, such that any two nearest-neighbouring state-variables $h_i$ and $h_{i+1}$ satisfy 
$h_{\, i} + h_{\, i + 1} \leq 1$, the hard-square model is the first example of a solved 
restricted-solid-on-solid model. 

\subsection{The restricted solid-on-solid models.} 
\label{work.andrews.baxter.forrester.01}
In \cite{andrews.baxter.forrester}, Andrews, Baxter and Forrester introduced an infinite series of statistical 
mechanical models with Boltzmann weights that satisfy the face-version of the Yang-Baxter equations, and that 
generalize the hard-square model. These are the restricted solid-on-solid models $\cL_{\, m, \, m+1}$, 
$m \in \ll 3, 4, \cdots \rr$
\footnote{\,
Also called the Andrews-Baxter-Forrester models. The hard-square model is $\cL_{\, 4, \, 5}$.
}. 

\subsubsection{The state variables}
\begin{figure}
\begin{tikzpicture}[scale=1.0]

\draw [thin] (-2.025,-3.5)--(-2.025, 1.5);
\draw [thin] (-1.025,-3.5)--(-1.025, 1.5);
\draw [thin] (-0.025,-3.5)--(-0.025, 1.5);
\draw [thin] ( 1.025,-3.5)--( 1.025, 1.5);
\draw [thin] ( 2.025,-3.5)--( 2.025, 1.5);

\draw [thin] (-2.500,-3.0)--(02.500,-3.0);
\draw [thin] (-2.500,-2.0)--(02.500,-2.0);
\draw [thin] (-2.500,-1.0)--(02.500,-1.0);
\draw [thin] (-2.500, 0.0)--(02.500, 0.0);
\draw [thin] (-2.500, 1.0)--(02.500, 1.0);

\foreach \x in {3,...,7}
{
\draw [fill=black!50] (\x-5, 1) circle (0.12);
\draw [fill=black!50] (\x-5, 0) circle (0.12);
\draw [fill=black!50] (\x-5,-1) circle (0.12);
\draw [fill=black!50] (\x-5,-2) circle (0.12);
\draw [fill=black!50] (\x-5,-3) circle (0.12);
}

\node at (-2.25,-3.25) {$\textswab{4}$};
\node at (-1.25,-3.25) {$\textswab{3}$};
\node at (-0.25,-3.25) {$\textswab{4}$};
\node at ( 0.75,-3.25) {$\textswab{3}$};
\node at ( 1.75,-3.25) {$\textswab{2}$};

\node at (-2.25,-2.25) {$\textswab{3}$};
\node at (-1.25,-2.25) {$\textswab{2}$};
\node at (-0.25,-2.25) {$\textswab{3}$};
\node at ( 0.75,-2.25) {$\textswab{2}$};
\node at ( 1.75,-2.25) {$\textswab{1}$};

\node at (-2.25,-1.25) {$\textswab{4}$};
\node at (-1.25,-1.25) {$\textswab{3}$};
\node at (-0.25,-1.25) {$\textswab{2}$};
\node at ( 0.75,-1.25) {$\textswab{1}$};
\node at ( 1.75,-1.25) {$\textswab{2}$};

\node at (-2.25,-0.25) {$\textswab{3}$};
\node at (-1.25,-0.25) {$\textswab{2}$};
\node at (-0.25,-0.25) {$\textswab{1}$};
\node at ( 0.75,-0.25) {$\textswab{2}$};
\node at ( 1.75,-0.25) {$\textswab{1}$};

\node at (-2.25, 0.75) {$\textswab{2}$};
\node at (-1.25, 0.75) {$\textswab{1}$};
\node at (-0.25, 0.75) {$\textswab{2}$};
\node at ( 0.75, 0.75) {$\textswab{3}$};
\node at ( 1.75, 0.75) {$\textswab{2}$};

\end{tikzpicture}
\caption{\textit{A configuration in the restricted solid-on-solid model 
$\cL_{\, 4, 5}$, where the state-variables take values in 
$\{\textswab{1}, \textswab{2}, \textswab{3}, \textswab{4}\}$.}}
\label{figure.restricted.solid.on.solid}
\end{figure}
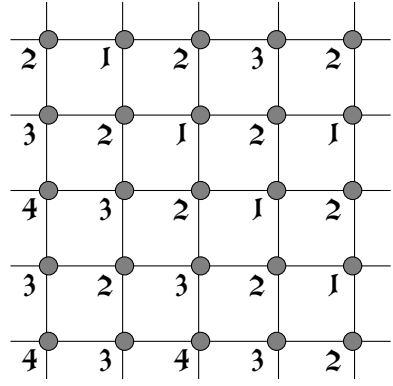

Following \cite{andrews.baxter.forrester}, a restricted solid-on-solid model is a statistical mechanical system 
of interacting state-variables that are also called \textit{\lq heights\rq}. Each state-variable $h$ lives at 
the intersection of a horizontal line and a vertical line in the lattice, and takes values in a finite set of 
integers $h \in \ll 1, 2, \cdots, m \rr $, where $m \in \NN$ characterizes the model. Adjacent state-variables differ 
by $\pm 1$. A configuration in the restricted solid-on-solid model $\cL_{\, 4, 5}$ is shown in Figure 
{\bf \ref{figure.restricted.solid.on.solid}}. 

\subsubsection{Remark} 
The hard-square model with occupancies $\ll 0, 1 \rr$, 
such that nearest-neighbouring state-variables satisfy $h_i + h_{i+1} \leq 1$, is equivalent to 
the $\cL_{\, 4, 5}$ model with occupancies $\ll 1, 2, 3, 4 \rr$,
such that nearest-neighbouring state-variables satisfy $h_i + h_{i+1} = \pm 1$, because 
any linear sequence of the first type can be mapped to a sequence of the second type, up to an involution, 
For example, 
$\ll 0, 0, 0, 1, 0, 1, 0, 0, 1, 0, 0 \rr$ maps to  
$\ll 2, 3, 2, 1, 2, 1, 2, 3, 4, 3, 2 \rr$, and also to
$\ll 3, 2, 3, 4, 3, 4, 3, 2, 1, 2, 3 \rr$, 
every configuration in the hard-square model corresponds to two configurations in the $\cL_{\, 4, 5}$ model, 
and the partition function of the hard-square model is related to that of the $\cL_{\, 4, 5}$ model as 
$Z_{\, hard \ hexagon} = \frac12 \ Z_{\, 4, 5}$ 

\subsubsection{Notation}
Given the elliptic theta functions
\footnote{\,
Section \textbf{8.18} of \cite{gradshteyn.ryzhik}
}, 

\begin{multline}
\theta_{\, 1}  \ll \eta x, p \rr  =   
2 \, p^{\, \frac14} \, sin \ll \eta \, x \rr \, 
\prod_{n=1}^{\, \infty} \ll 1 - 2 \, cos \ll 2 \, \eta \, x \rr \, p^{\, 2 n} \, + p^{\, 4 n} \rr
\ll 1 - p^{\, 2 n} \rr,
\\
\theta_{\, 4} \ll \eta x, p \rr =   
\prod_{n=1}^{\, \infty} \ll 1 - 2 \, cos \ll 2 \, \eta \, x \rr \, p^{\, 2n-1} \, + p^{\, 2n-2} \rr
\ll 1 - p^{\, 2 n} \rr,
\end{multline}

\noindent we define the bracket $[x, p]$, or simply $[x]$ as short-hand notation for the product,  

\begin{multline}
[x, p] = [x] = \theta_{\, 1}  \ll \eta x, p \rr \theta_{\, 4} \ll \eta x, p \rr =  
\\
2 \, p^{\, \frac14} \, sin \ll \eta \, x \rr \, 
\prod_{n=1}^{\infty} \ll 1 - 2 \, cos \ll 2 \, \eta \, x \rr \, p^{\, n} \, + p^{\, 2n} \rr
\ll 1 - p^{\, 2 n} \rr^{\, 2}
\label{bracket}    
\end{multline}

\noindent In the sequel, $\eta = \frac{\pi}{m+1}$ is the crossing parameter of the model 
$\cL_{\, m, m+1}$, 
$p = e^{- \, 2 \, \pi / \tau}$ is the nome of the elliptic theta function, and modular 
parameter $\tau$ measures the departure of the model from criticality, in the sense that,
in the limit $\tau \rightarrow 0$, $p \rightarrow 0$, the Boltzmann weights become 
trigonometric, and the model becomes critical 
\footnote{\, 
We write the modular parameter of the nome $p$ as $\tau^{\, \prime} = 1/ \tau$, instead 
of $\tau$, for later convenience.
}.

\subsubsection{Remark} 
\label{remark.01}
It is useful to note that 
$[x, p^2] = 
\theta_{\, 1} \ll \eta x, p^2 \rr \, \theta_{\, 4} \ll \eta x, p^2 \rr = 
C \ll p \rr \ \theta_{\, 1} \ll \eta x, p \rr$, 
where $C [p]$ depends on $p$ only. 

\subsubsection{The Boltzmann weights}
\label{the.boltzmann.weights}
Each set of four state-variables       $  \ll h_{\, 1}, h_{\, 2}, h_{\, 3}, h_{\, 4}               \rr$ at the corners 
of a face is assigned a weight $w \ll h_{\, 1}, h_{\, 2}, h_{\, 3}, h_{\, 4} \, \vert \, u \rr$. 
There are six non-zero weights 
$w \ll h_{\, 1}, h_{\, 2}, h_{\, 3}, h_{\, 4} \, \vert \, u \rr$, such that adjacent state-variables 
differ by $\pm 1$, 

\begin{eqnarray}
w \ll h, h \pm 1, h \pm 2, h \pm 1 \, \vert \, u \rr & = & 
\frac{[u + 1]}{[1]}, 
\label{weight.01}
\\
w \ll h, h \pm 1, h,       h \mp 1 \, \vert \, u \rr & = & 
\frac{[h \pm 1] \, [u]}{[h] \, [1]},  
\label{weight.02}
\\
w \ll h, h \pm 1, h,       h \pm 1 \, \vert \, u \rr & = & 
\frac{[h \mp u]}{[h]}, 
\label{weight.03}
\end{eqnarray}

\noindent where $u$ is a spectral parameter, and the bracket $[x]$ is defined in (\ref{bracket}). 
All other weights are set to zero. The weights in (\ref{weight.01}--\ref{weight.03}) 
satisfy the interaction-\-round-\-a-\-face version of Yang-Baxter equations \cite{andrews.baxter.forrester}. 
Depending on the choice of the remaining parameters, $\cL_{\, m, m+1}$ can be in regime-I, -II, -III, or in 
regime-IV. We restrict our attention to regime-III, where $p, u \in \RR$, and,  

\begin{equation}
 0 < p < 1,   
\quad  
\textit{and}
\quad 
-1 < u < 0, 
\label{choice}
\end{equation}

\noindent which guarantees that the Boltzmann weights are real. For a complete discussion of the restricted solid-on-solid 
models, we refer to \cite{andrews.baxter.forrester}.

\subsubsection{Remark} 
\label{remark.02}
The choice $-1 < u < 0$ in (\ref{choice}) makes the weights $w \ll h, h \pm 1, h, h \mp 1 \, | \, u \rr$  
in (\ref{weight.02}) negative, but the signs of these weights can be changed without affecting the Yang-Baxter 
equations. This is because these weights appear either exactly once in each term in some Yang-Baxter equations, or an even 
number of times in each term in the remaining ones
\footnote{\, 
Equation (1.4.7) in \cite{andrews.baxter.forrester}
}. 

\subsubsection{Remark}
\label{remark.03}
From section \textbf{\ref{remark.01}}, the Boltzmann weights in 
(\ref{weight.01}--\ref{weight.03}) could be written in the same form but in terms of the theta functions 
$\theta_{\, 1} \ll x, p \rr$, as the factors $C \ll p \rr$ of section \textbf{\ref{remark.01}} cancel out. Both forms of the 
weights appear in the literature. 

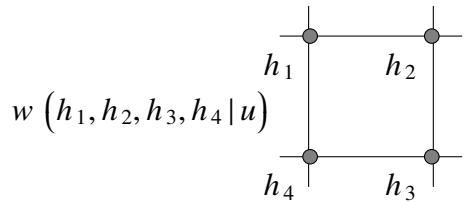
\begin{figure}
\begin{tikzpicture}[scale=0.8]

\draw [thin] (-1.025,-2.5)--(-1.025, 0.5);
\draw [thin] ( 1.025,-2.5)--( 1.025, 0.5);

\draw [thin] (-1.500,-2.0)--(01.500,-2.0);
\draw [thin] (-1.500, 0.0)--(01.500, 0.0);

\foreach \x in {4,6}
{
\draw [fill=black!50] (\x-5, 0) circle (0.12);
\draw [fill=black!50] (\x-5,-2) circle (0.12);
}

\node at (-3.75,-1.25) {$w \ll h_{\, 1}, h_{\, 2}, h_{\, 3}, h_{\, 4} \, \vert \, u \rr$};

\node at (-1.50,-0.50) {$h_{\, 1}$};
\node at ( 0.50,-0.50) {$h_{\, 2}$};
\node at ( 0.50,-2.50) {$h_{\, 3}$};
\node at (-1.50,-2.50) {$h_{\, 4}$};

\node at ( 3.75,-1.25) {\phantom{$w \ll h_{\, 1}, h_{\, 2}, h_{\, 3}, h_{\, 4} \, \vert \, u \rr$}};

\end{tikzpicture}
\caption{\textit{
The weight 
$w \ll h_{\, 1}, h_{\, 2}, h_{\, 3}, h_{\, 4} \, \vert \, u \rr$ associated to a face 
on the lattice with state-variables $\ll h_{\, 1}, h_{\, 2}, h_{\, 3}, h_{\, 4} \rr$ 
at the corners, and a spectral parameter $u$.
}}
\label{figure.faces.and.weights}
\end{figure}

\subsubsection{Remark}
We use the notation $\cL_{\, m, \, m+1}$ for the restricted solid-on-solid models considered in this work 
because these models flow at criticality to the unitary minimal conformal field theories $\cM_{\, m, \, m+1}$
\cite{friedan.qiu.shenker.01, friedan.qiu.shenker.02}, which form a subset of the minimal conformal field 
theories $\cM_{\, m, \, m^{\, \prime}}$, where $m$ and $m^{\, \prime}$ are co-prime positive integers
\cite{belavin.polyakov.zamolodchikov.01, belavin.polyakov.zamolodchikov.02}.

\subsection{Other restricted solid-on-solid models}

\subsubsection{The Forrester-Baxter models}
The $\cL_{\, m, \, m+1}$ models are obtained by setting the crossing parameter of the unrestricted solid-on-solid 
model $\eta = \pi / \ll m + 1 \rr$. In \cite{forrester.baxter}, Forrester and Baxter set $\eta = n \, \pi / \ll m + 1 \rr$,  
$n = 2, \cdots, m$, and obtained a more general series of restricted solid-on-solid models that flow non-unitary 
minimal conformal field theories. 

\subsubsection{Type-$D$ restricted solid-on-solid models} 
\label{work.pasquier}
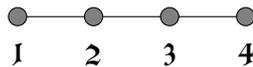
\begin{figure}
\begin{tikzpicture}[scale=1.0]

\draw [thin] (-2.000,-1.0)--(01.000,-1.0);

\foreach \x in {3,...,6}
{
\draw [fill=black!50] (\x-5,-1) circle (0.12);
}

\node at (-2.00,-1.50) {$\textswab{1}$};
\node at (-1.00,-1.50) {$\textswab{2}$};
\node at (-0.00,-1.50) {$\textswab{3}$};
\node at ( 1.00,-1.50) {$\textswab{4}$}; 

\end{tikzpicture}
\caption{\textit{The Dynkin diagram $A_{\, 4}$ that corresponds to the restricted solid-on-solid 
model $\cL_{\, 4, 5}$, where the state-variables take values in 
$\ll \textswab{1}, \textswab{2}, \textswab{3}, \textswab{4}\rr$.}}
\label{figure.A.4}
\end{figure}

The off-critical solid-on-solid models of Andrews, Baxter and Forrester are related to $A_{\, m}$ 
Lie algebras, starting from the fact that the allowed state-variables and their adjacencies correspond to 
the nodes of an $A_{\, m}$, $m = 3, 4, \cdots$, Dynkin diagram, and the height state variables 
take values in $\ll 1, 2, \cdots, m\rr$. See Figure \textbf{ \ref{figure.A.4}}. 
It is common to refer to the models of Andrews \textit{et al.} as type-$A$ models. 
In \cite{pasquier.01}, Pasquier introduced a class of off-critical solid-on-solid models whose 
allowed state-variables and their adjacencies correspond to the nodes of a $D_{\, m}$, $m = 4, 5 \cdots$,
Dynkin diagram, and the height state variables take values in $\ll 1, \cdots, m-1, \overline{m-1}\, 
\rr$, and in \cite{pasquier.02, pasquier.03}, solved these models and computed their local 
height probabilities. It is common to refer to the models of Pasquier as type-$D$ models
\footnote{\,
There are many more classes of generalized restricted solid-on-solid models, including 
the critical restricted solid-on-solid models based $E_6$, $E_7$ and $E_8$ Dynkin diagrams, 
also due to Pasquier \cite{pasquier.01}, but we do not need to recall these in the present 
work. For further references, we refer the reader to \cite{itzykson.saleur.zuber.book}
}${}^{,}$
\footnote{\,
In another contribution \cite{pasquier.04}, of relevance to the present work, Pasquier 
extended the Coulomb gas representation \cite{nienhuis.01, nienhuis.02} to the critical 
restricted solid-on-solid models. 
}. 
See Figure \textbf{ \ref{figure.D.4}}. 

\begin{figure}
\begin{tikzpicture}[scale=1.0]

\draw [thin] (-2.000,-1.0)--(00.000,-1.0);

\draw [thin] (00.000,-1.0)--( 0.750, -0.25);
\draw [thin] (00.000,-1.0)--( 0.750, -1.75);

\foreach \x in {3,...,5}
{
\draw [fill=black!50] (\x-5,-1) circle (0.12);
}

\draw [fill=black!50] (0.750, -0.25) circle (0.12);
\draw [fill=black!50] (0.750, -1.75) circle (0.12);

\node at (-2.00,-1.50) {$\textswab{1}$};
\node at (-1.00,-1.50) {$\textswab{2}$};
\node at (-0.00,-1.50) {$\textswab{3}$};

\node at ( 0.75, -0.75) {    $\textswab{4}$};
\node at ( 0.75, -2.25) {\ol{$\textswab{4}$}};

\end{tikzpicture}
\caption{\textit{The Dynkin diagram $D_{\, 5}$ that corresponds to the restricted solid-on-solid 
model $\cL^{\, D}_{\, 5, 6}$, where the state-variables take values in 
$\ll \textswab{1}, \textswab{2}, \textswab{3}, \textswab{4}, \ol{\textswab{4}} \rr$.}}
\label{figure.D.4}
\end{figure}

\begin{figure}
\begin{tikzpicture}[scale=1.0]

\draw [fill=black!00] (0,0) circle (1.25);

\draw [fill=black!50] (  1.25,0) circle (0.12);
\draw [fill=black!50] ( -1.25,0) circle (0.12);
\draw [fill=black!50] (     0, 1.25) circle (0.12);
\draw [fill=black!50] (     0,-1.25) circle (0.12);

\node at ( 1.50, -0.25) {$\textswab{1}$};
\node at (    0, -1.75) {$\textswab{2}$}; 
\node at (-1.00, -0.25) {$\textswab{3}$};
\node at (    0,  0.75) {$\textswab{4}$};

\end{tikzpicture}
\caption{\textit{The Dynkin diagram of the affine Lie algebra $\widehat{A}_{\, 4}$ that corresponds 
to the cyclic solid-on-solid model $\cL^{\, cyc}_{\, 4, 5}$, where the state-variables take values 
in $\ll \textswab{1}, \textswab{2}, \textswab{3}, \textswab{4} \rr$, and the height variables 
$\textswab{1}$ and $\textswab{4}$ are regarded as nearest-neighbours.}}
\label{figure.cyclic.A.4}
\end{figure}
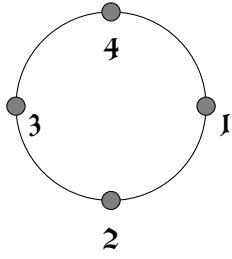

\subsubsection{The cyclic solid-on-solid models} 
In \cite{kuniba.yajima.01, kuniba.yajima.02}, Kuniba and Yajima studied cyclic solid-on-solid models
where the height variables take values in the nodes of affine $\widehat{A_m}$ Dynkin diagrams, see 
Figure \textbf{\ref{figure.cyclic.A.4}}, and used the corner transfer matrix method to compute their 
local height probabilities. 
In \cite{kuniba.yajima.01, kuniba.yajima.02}, Kuniba and Yajima also study solid-on-solid models based 
on $\widehat{D_m}$ Dynkin diagrams, but these results will not be referred to in the sequel.
In \cite{pearce.seaton.01, pearce.seaton.02} studied the cyclic solid-on-solid models based on 
$\widehat{A_m}$ Dynkin diagrams, in the presence of an extra parameter that acts as a simultaneous 
additive shift of the height variables.   

\section{From corner transfer matrices to local height probabilities}
\label{section.corner.transfer.matrices.local.height.probabilities}
\textit{
We recall basic facts related to Baxter's corner transfer matrix method, and define the local height probabilities. 
}
\subsection{The corner transfer matrix method} The corner transfer matrix method 
originates in Baxter's work on 
the monomer-dimer problem \cite{baxter.1968}, and was put in its current graphical form in 
\cite{baxter.1978}. For an introduction to the method, and an overview of its development, 
see \cite{baxter.review, baxter.book}. 
In \cite{baxter.1980, baxter.1981}, Baxter used this method to compute the order parameters 
in the hard-hexagon model, and made an unexpected connection with the Rogers-Ramanujan 
identities \cite{andrews.1981}. This development led to the development of the restricted 
solid-on-solid models as an infinite series that includes the hard-hexagon model as a special 
case \cite{andrews.baxter.forrester}, and the corner transfer matrix method was used to compute 
the off-critical local height probabilities. 

\subsubsection{Quadrants, boundaries, initial and final states}
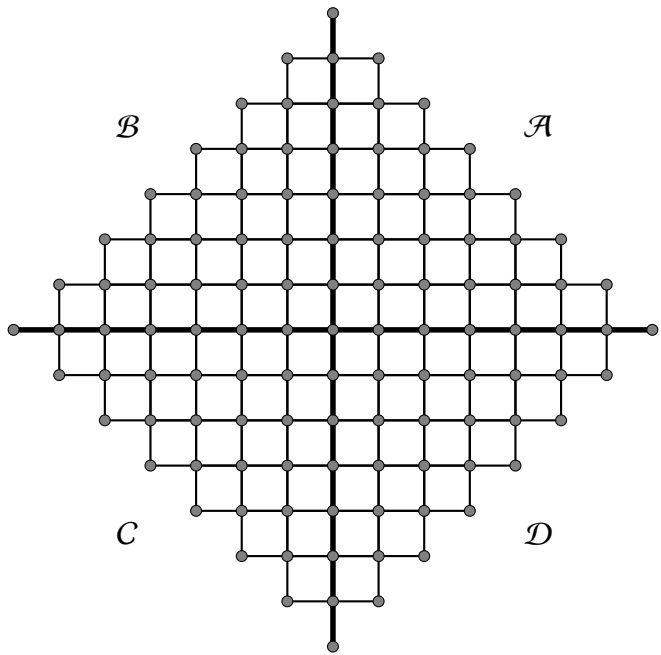
\begin{figure}
\begin{tikzpicture}[scale=.6]


\draw [thick] (0, 5) rectangle (1,6);

\draw [thick] (0, 4) rectangle (1,5);
\draw [thick] (1, 4) rectangle (2,5);

\draw [thick] (0, 3) rectangle (1,4);
\draw [thick] (1, 3) rectangle (2,4);
\draw [thick] (2, 3) rectangle (3,4);

\draw [thick] (0, 2) rectangle (1,3);
\draw [thick] (1, 2) rectangle (2,3);
\draw [thick] (2, 2) rectangle (3,3);
\draw [thick] (3, 2) rectangle (4,3);

\draw [thick] (0, 1) rectangle (1,2);
\draw [thick] (1, 1) rectangle (2,2);
\draw [thick] (2, 1) rectangle (3,2);
\draw [thick] (3, 1) rectangle (4,2);
\draw [thick] (4, 1) rectangle (5,2);

\draw [thick] (0, 0) rectangle (1,1);
\draw [thick] (1, 0) rectangle (2,1);
\draw [thick] (2, 0) rectangle (3,1);
\draw [thick] (3, 0) rectangle (4,1);
\draw [thick] (4, 0) rectangle (5,1);
\draw [thick] (5, 0) rectangle (6,1);

\draw [very thick] (-0.025,-7.0)--(-0.025,7.0);
\draw [very thick] ( 0.025,-7.0)--( 0.025,7.0);

\draw [very thick] (-7.0, -0.025)--(7.0, -0.025);
\draw [very thick] (-7.0,  0.025)--(7.0,  0.025);

\draw [thick] (0, -1) rectangle (1,0);
\draw [thick] (1, -1) rectangle (2,0);
\draw [thick] (2, -1) rectangle (3,0);
\draw [thick] (3, -1) rectangle (4,0);
\draw [thick] (4, -1) rectangle (5,0);
\draw [thick] (5, -1) rectangle (6,0);

\draw [thick] (0, -2) rectangle (1,-1);
\draw [thick] (1, -2) rectangle (2,-1);
\draw [thick] (2, -2) rectangle (3,-1);
\draw [thick] (3, -2) rectangle (4,-1);
\draw [thick] (4, -2) rectangle (5,-1);

\draw [thick] (0, -3) rectangle (1,-2);
\draw [thick] (1, -3) rectangle (2,-2);
\draw [thick] (2, -3) rectangle (3,-2);
\draw [thick] (3, -3) rectangle (4,-2);

\draw [thick] (0, -4) rectangle (1,-3);
\draw [thick] (1, -4) rectangle (2,-3);
\draw [thick] (2, -4) rectangle (3,-3);

\draw [thick] (0, -5) rectangle (1,-4);
\draw [thick] (1, -5) rectangle (2,-4);

\draw [thick] (0, -6) rectangle (1,-5);


\draw [thick] ( 0, 5) rectangle (-1,6);

\draw [thick] ( 0, 4) rectangle (-1,5);
\draw [thick] (-1, 4) rectangle (-2,5);

\draw [thick] ( 0, 3) rectangle (-1,4);
\draw [thick] (-1, 3) rectangle (-2,4);
\draw [thick] (-2, 3) rectangle (-3,4);

\draw [thick] ( 0, 2) rectangle (-1,3);
\draw [thick] (-1, 2) rectangle (-2,3);
\draw [thick] (-2, 2) rectangle (-3,3);
\draw [thick] (-3, 2) rectangle (-4,3);

\draw [thick] ( 0, 1) rectangle (-1,2);
\draw [thick] (-1, 1) rectangle (-2,2);
\draw [thick] (-2, 1) rectangle (-3,2);
\draw [thick] (-3, 1) rectangle (-4,2);
\draw [thick] (-4, 1) rectangle (-5,2);

\draw [thick] ( 0, 0) rectangle (-1,1);
\draw [thick] (-1, 0) rectangle (-2,1);
\draw [thick] (-2, 0) rectangle (-3,1);
\draw [thick] (-3, 0) rectangle (-4,1);
\draw [thick] (-4, 0) rectangle (-5,1);
\draw [thick] (-5, 0) rectangle (-6,1);

\draw [thick] ( 0, -1) rectangle (-1,0);
\draw [thick] (-1, -1) rectangle (-2,0);
\draw [thick] (-2, -1) rectangle (-3,0);
\draw [thick] (-3, -1) rectangle (-4,0);
\draw [thick] (-4, -1) rectangle (-5,0);
\draw [thick] (-5, -1) rectangle (-6,0);

\draw [thick] ( 0, -2) rectangle (-1,-1);
\draw [thick] (-1, -2) rectangle (-2,-1);
\draw [thick] (-2, -2) rectangle (-3,-1);
\draw [thick] (-3, -2) rectangle (-4,-1);
\draw [thick] (-4, -2) rectangle (-5,-1);

\draw [thick] ( 0, -3) rectangle (-1,-2);
\draw [thick] (-1, -3) rectangle (-2,-2);
\draw [thick] (-2, -3) rectangle (-3,-2);
\draw [thick] (-3, -3) rectangle (-4,-2);

\draw [thick] ( 0, -4) rectangle (-1,-3);
\draw [thick] (-1, -4) rectangle (-2,-3);
\draw [thick] (-2, -4) rectangle (-3,-3);

\draw [thick] ( 0, -5) rectangle (-1,-4);
\draw [thick] (-1, -5) rectangle (-2,-4);

\draw [thick] ( 0, -6) rectangle (-1,-5);

\foreach \x in {0,...,14}
{
\draw [fill=black!50] (\x-7, 0) circle (0.12);
}

\foreach \x in {0,...,12}
{
\draw [fill=black!50] (\x-6, 1) circle (0.12);
\draw [fill=black!50] (\x-6,-1) circle (0.12);
}

\foreach \x in {0,...,10}
{
\draw [fill=black!50] (\x-5, 2) circle (0.12);
\draw [fill=black!50] (\x-5,-2) circle (0.12);
}

\foreach \x in {0,...,8}
{
\draw [fill=black!50] (\x-4, 3) circle (0.12);
\draw [fill=black!50] (\x-4,-3) circle (0.12);
}

\foreach \x in {0,...,6}
{
\draw [fill=black!50] (\x-3, 4) circle (0.12);
\draw [fill=black!50] (\x-3,-4) circle (0.12);
}

\foreach \x in {0,...,4}
{
\draw [fill=black!50] (\x-2, 5) circle (0.12);
\draw [fill=black!50] (\x-2,-5) circle (0.12);
}

\foreach \x in {0,...,2}
{
\draw [fill=black!50] (\x-1, 6) circle (0.12);
\draw [fill=black!50] (\x-1,-6) circle (0.12);
}

\draw [fill=black!50] (0,  7) circle (0.12);
\draw [fill=black!50] (0, -7) circle (0.12);

\node at ( 4.5, 4.5) {$\cA$};
\node at (-4.5, 4.5) {$\cB$};
\node at (-4.5,-4.5) {$\cC$};
\node at ( 4.5,-4.5) {$\cD$};

\end{tikzpicture}
\caption{\textit{
The splitting of the lattice into four quadrants. We assume that the corner transfer matrices act 
counterclockwise. The corner transfer matrix $\cA$ acts on 
the horizontal line of vertices that extends from the center to the right-most vertex, and generates
the vertical   line of vertices that extends from the center to the top site. This \lq transition\rq\ 
is weighted by the partition function of this quadrant. $\cB$, $\cA$ and $\cD$ act analogously.
}}
\label{figure.4.quadrants}
\end{figure}

Following \cite{baxter.book, andrews.baxter.forrester}, we divide the square lattice into four quadrants, 
as in Figure \textbf{\ref{figure.4.quadrants}}. Each quadrant is bounded by 
an inner boundary, which is the vertex at the center of the lattice that carries a state-variable denoted by 
$a \in \ll 1, \cdots, m \rr$, 
an outer boundary that consists of a sequence of vertices such that adjacent vertices carry state-variables 
denoted by $\ll b, b \pm 1 \rr \in \ll 1, \cdots, n \rr$, 
an initial half-line that one can regard as an initial state, and 
a  final   half-line that one can regard as a  final state. 

\subsubsection{Corner transfer matrices as operator}
A corner transfer matrix acts on the height variables on the initial half-line, and generates the height 
variables on the final half-line, such that the state-variables on the inner and outer boundaries are preserved. 
This action can be regarded as a transition from an initial to a final state, with a transition probability 
that can be computed in terms of the Boltzmann weights of the model. We denote 
the state-variables on the initial half-line and on the final half-line by, 

\begin{equation} 
\textbf{ h}        = \ll h_0,        h_{\, 1} ,        \cdots, h_{\, L},        h_{L+1}         \rr, 
\quad
{\bf  h^{\, \prime}}    = \ll h_0^{\, \prime}, h_{\, 1}^{\, \prime}, \cdots, h_{\, L}^{\, \prime}, h_{L+1}^{\, \prime}  \rr,
\end{equation}

\noindent respectively, where $h_0 = h_0^{\, \prime}$ is the state-variable at the center of the lattice
\footnote{\,
In \cite{andrews.baxter.forrester}, the state-variable at the center of the lattice is $h_{\, 1}$. 
In the present note, we use $h_0$. 
}. 
An element of the corner transfer matrix element is the partition function of a quadrant, for fixed ${\bf  h}$ and 
${\bf  h^{\, \prime}}$, that is, the sum over all allowed weighted configurations. 
The weight of an allowed configuration is the product of the weights of the faces of the configurations. We denote 
the corner transfer matrices of the four quadrants by $\cA, \cB, \cC$, and $\cD$. We refer to 
\cite{baxter.book, andrews.baxter.forrester} for the definition of the corner transfer matrix in terms of the Boltzmann 
weights. 

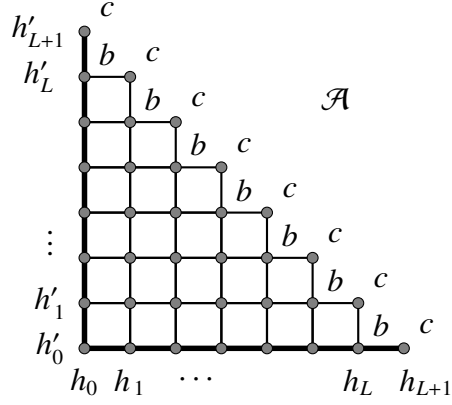
\begin{figure}
\begin{tikzpicture}[scale=.6]


\draw [thick] (0, 5) rectangle (1,6);

\draw [thick] (0, 4) rectangle (1,5);
\draw [thick] (1, 4) rectangle (2,5);

\draw [thick] (0, 3) rectangle (1,4);
\draw [thick] (1, 3) rectangle (2,4);
\draw [thick] (2, 3) rectangle (3,4);

\draw [thick] (0, 2) rectangle (1,3);
\draw [thick] (1, 2) rectangle (2,3);
\draw [thick] (2, 2) rectangle (3,3);
\draw [thick] (3, 2) rectangle (4,3);

\draw [thick] (0, 1) rectangle (1,2);
\draw [thick] (1, 1) rectangle (2,2);
\draw [thick] (2, 1) rectangle (3,2);
\draw [thick] (3, 1) rectangle (4,2);
\draw [thick] (4, 1) rectangle (5,2);

\draw [thick] (0, 0) rectangle (1,1);
\draw [thick] (1, 0) rectangle (2,1);
\draw [thick] (2, 0) rectangle (3,1);
\draw [thick] (3, 0) rectangle (4,1);
\draw [thick] (4, 0) rectangle (5,1);
\draw [thick] (5, 0) rectangle (6,1);

\draw [very thick] (-0.025,-0.0)--(-0.025,7.0);
\draw [very thick] ( 0.025,-0.0)--( 0.025,7.0);

\draw [very thick] (-0.0, -0.025)--(7.0, -0.025);
\draw [very thick] (-0.0,  0.025)--(7.0,  0.025);


\foreach \x in {7,...,14}
{
\draw [fill=black!50] (\x-7, 0) circle (0.12);
}

\foreach \x in {6,...,12}
{
\draw [fill=black!50] (\x-6, 1) circle (0.12);
}

\foreach \x in {5,...,10}
{
\draw [fill=black!50] (\x-5, 2) circle (0.12);
}

\foreach \x in {4,...,8}
{
\draw [fill=black!50] (\x-4, 3) circle (0.12);
}

\foreach \x in {3,...,6}
{
\draw [fill=black!50] (\x-3, 4) circle (0.12);
}

\foreach \x in {2,...,4}
{
\draw [fill=black!50] (\x-2, 5) circle (0.12);
}

\foreach \x in {1,...,2}
{
\draw [fill=black!50] (\x-1, 6) circle (0.12);
}

\draw [fill=black!50] (0,  7) circle (0.12);

\node at (5.5, 5.5) {$\cA$};

\node at (0.0, -0.75) {$h_0$};
\node at (1.0, -0.75) {$h_{\, 1}$};
\node at (2.5, -0.75) {$\cdots$};
\node at (6.0, -0.75) {$h_{L  }$};
\node at (7.5, -0.75) {$h_{L+1}$};

\node at (-0.75, 0.00) {$h_0^\prime$};
\node at (-0.75, 1.00) {$h_{\, 1}^\prime$};
\node at (-0.75, 2.50) {$\vdots$};
\node at (-1.00, 6.00) {$h_{L  }^\prime$};
\node at (-1.00, 7.00) {$h_{L+1}^\prime$};

\node at ( 0.50, 6.50) {$b$};
\node at ( 1.50, 5.50) {$b$};
\node at ( 2.50, 4.50) {$b$};
\node at ( 3.50, 3.50) {$b$};
\node at ( 4.50, 2.50) {$b$};
\node at ( 5.50, 1.50) {$b$};
\node at ( 6.50, 0.50) {$b$};

\node at ( 0.50, 7.50) {$c$};
\node at ( 1.50, 6.50) {$c$};
\node at ( 2.50, 5.50) {$c$};
\node at ( 3.50, 4.50) {$c$};
\node at ( 4.50, 3.50) {$c$};
\node at ( 5.50, 2.50) {$c$};
\node at ( 6.50, 1.50) {$c$};
\node at ( 7.50, 0.50) {$c$};

\end{tikzpicture}
\caption{\textit{ 
The $\cA$ quadrant, with state-variables  
$h_i$,             $i = 0, 1, \cdots, L+1$, on the initial half-line/state, and
$h_i^{\, \prime}$, $i = 0, 1, \cdots, L+1$, on the final   half-line/state, such that 
the state-variable  on the inner boundary $h_0 = h_0^{\, \prime} = a$, and 
the state variables on the outer boundaries alternate as $\ll b, c = b \pm 1 \rr$, and
$a, b, c \in \ll 1, 2, \cdots, m \rr$. 
}}
\label{figure.1.quadrant}
\end{figure}

\subsubsection{The partition function with fixed outer boundary conditions}
Let $\cL_{m, m + 1}$ be a regime-III restricted solid-on-solid model as in \cite{andrews.baxter.forrester}. 
The off-critical partition function on a plane such that the state-variables on the outer boundary are fixed alternately at 
the ground-state pair $\ll b, b \pm 1 \rr$ can be written in terms of corner transfer matrices as, 

\begin{equation}
Z\ll b, b \pm 1 \, \vert \, p \rr = 
\textit{Trace} \ll \, \cA\, \cB\, \cC\, \cD\,  \rr
\label{partition.function}
\end{equation}

\noindent where the trace over the product of corner transfer matrices in (\ref{partition.function}) indicates 
identifying the state-variables on the common half-diagonals, then summing over all state-variables that are not fixed by the outer boundary 
conditions, in all possible ways. 

\subsection{The local height probabilities}
\subsubsection{Fixing the height at the inner boundary}
The partition function $Z \ll a, b, b \pm 1 \, \vert \, p \rr$ such that the state-variable $h_{\, 0}$, at the center 
of the lattice, is fixed, $h_{\, 0} = a$, and the state-variables on the outer boundary are fixed alternately at the 
ground-state pair $\ll b, b \pm 1 \rr$, can be computed using corner transfer matrices by inserting the diagonal matrix 
$\cS_a$, where,  

\begin{equation}
\ll \cS_a \, \rr_{h, \, h^{\, \prime} } = 
\delta\ll h_0, \, a \rr \delta \ll \h, \, \h^{\, \prime}  \rr, 
\quad
\delta \ll \h, \, \h^{\, \prime}  \rr 
\equiv 
\prod_{i=0}^{L+1} \delta\ll h_i, \, h^{\, \prime}_i \rr, 
\end{equation}

\noindent in (\ref{partition.function}), to obtain, 

\begin{equation}
Z \ll a, b, \, b \pm 1 \, \vert \, p \rr = \textit{Trace} \ll \cS_a \, \cA \, \cB \, \cC \, \cD \rr, 
\end{equation}

\noindent where $Z \ll a, b, b \pm 1 \, \vert \, p \rr$ is the two-dimensional configuration sum of the system with 
fixed boundary conditions both at the center and the boundary. 

\subsection{Normalization to obtain probabilities}
The probability $P\ll a, b, b \pm 1 \, \vert \, p \rr$ that the height variable at the center of the lattice is $a$, while 
the state-variables on the boundary are $\ll b, b \pm 1 \rr $, is obtained by normalizing $Z \ll a, b, b \pm 1 \, \vert \, p \rr$ 
by $\sum_{a^{\, \prime}} Z\ll a^{\, \prime}, b, b \pm 1 \, \vert \, p \rr$, $a^{\, \prime} = 1, \cdots, m$, to obtain, 

\begin{equation}
P \ll a, b, b \pm 1 \, \vert \, p \rr  =
\frac{
\textit{Trace} \ll \cS_{\, a} \, \cA\, \cB\, \cC\, \cD\, \rr
}
{
\textit{Trace} \ll \, \cA\, \cB\, \cC\, \cD\, \rr
}, 
\quad 
\sum_{\, a^{\, \prime}} P \ll a^{\, \prime}, b,  b \pm 1 \, \vert \, p \rr = 1, 
\label{local.height.probability.01}
\end{equation}

\noindent and $P \ll a, b,  b \pm 1 \, \vert \, p \rr$ is indeed a probability.

\section{From local height probabilities to modular functions}
\label{section.planar.geometry}
\textit{We outline the evaluation of the local height probabilities on a plane in six steps, in preparation 
for section \textbf{\ref{section.spiral.geometry}}, where these steps carry over with minimal modification to 
the $4N$-quadrant case.
}

\subsection{Evaluating the local height probabilities} 
\label{evaluating.the.local.height.probabilities}
In Appendix \textbf{A} of \cite{andrews.baxter.forrester}, Andrews \textit{et al.} use the corner transfer matrix 
method to evaluate the off-critical local height probabilities, $P \ll a, \,  b,\, b \pm 1 \, \vert \, p \rr$, in 
a planar geometry as functions of the Boltzmann weights, which in turn are functions of elliptic theta functions 
of nome $p$. They do this essentially as follows.

\subsubsection{Step 1. Factorize the corner transfer matrices} 
The Yang-Baxter equations are used, in the thermodynamic limit, to factorize the corner transfer matrices in the form, 

\begin{multline}
\cA \ll u\rr = \cQ_{\, 1} \, \cM_{\, 1} e^{ \, u \, \cH} \, \cQ_{\, 2}^{-1},  \quad 
\cB \ll u\rr = \cQ_{\, 2} \, \cM_{\, 2} e^{-\, u \, \cH} \, \cQ_{\, 3}^{-1}, \\ 
\cC \ll u\rr = \cQ_{\, 3} \, \cM_{\, 3} e^{ \, u \, \cH} \, \cQ_{\, 4}^{-1},  \quad 
\cD \ll u\rr = \cQ_{\, 4} \, \cM_{\, 4} e^{-\, u \, \cH} \, \cQ_{\, 1}^{-1} 
\label{simplification.01}
\end{multline}

\noindent where the matrices $\cH, \cM_{\, 1}, \cdots, \cM_{\, 4}$, and $\cQ_{\, 1}, \cdots, \cQ_{\, 4}$ are independent of the spectral 
parameter $u$. This is a major simplification, as the dependence of each corner transfer matrix on the spectral 
parameter $u$ is now localized in an overall multiplicative factor in an exponent. The Yang-Baxter equations are also 
used to show that $\cH$ and $\cM_{\, 1}, \cdots, \cM_{\, 4}$ are diagonal, which is another major simplification that allows us 
to use (\ref{simplification.01}) in (\ref{local.height.probability.01}) to obtain,  

\begin{equation}
P \ll a, b, b \pm 1 \, \vert \, p \rr = 
\frac{
\textit{Trace}  \ll \cS_a \, \cM_{\, 1} \, \cM_{\, 2} \, \cM_{\, 3} \, \cM_{\, 4} \rr}{
\textit{Trace}  \ll          \cM_{\, 1} \, \cM_{\, 2} \, \cM_{\, 3} \, \cM_{\, 4} \rr}, 
\label{local.height.probability.02}
\end{equation}

\noindent and find that $P \ll a, b, b \pm 1 \, \vert \, p \rr$ is independent of the spectral parameter $u$, which is 
a magnificent result, in the sense that it is enormously simpler than the expression that we started with. 
\subsubsection{Step 2. Reduce the computation of $P \ll a, b, b \pm 1 \, \vert \, p \rr$ to a computation of $\cH$}
To evaluate the matrix products in (\ref{local.height.probability.02}), 
Andrews \textit{et al.} show that, 

\begin{equation}
\cA \ll  \eta \rr = \cC \ll  \eta \rr = \cI, \quad 
\cB \ll -\eta \rr = \cD \ll -\eta \rr = \cR_{\, a}, 
\label{simplified.corner.transfer.matrices}
\end{equation}

\noindent where $\eta$ is the crossing parameter, and $\cR_{\, a}$ is the diagonal matrix
\footnote{\, 
$\cR_{\, a}$ in this work is $\cR_{\, 1}$ in \cite{andrews.baxter.forrester}.
},  

\begin{equation}
\ll \cR_{\, a} \rr_{h, \, h^{\, \prime} } = [a]^{1/2} \, \delta \ll \h, \, \h^{\, \prime} \rr, 
\end{equation}

\noindent the bracket $[a]$ is defined in (\ref{bracket}), and $a$ is the fixed state-variable at the site at 
the centre of the lattice. From (\ref{simplification.01}) and (\ref{simplified.corner.transfer.matrices}), 
and the fact that $\cQ_i$, $i=1, \cdots, 4$, and $\cH$ are diagonal, we obtain,

\begin{equation}
\cA \ll \eta \rr  \, \cB \ll - \eta \rr  \, \cC \ll \eta \rr  \, \cD \ll - \eta \rr  = 
\ll \cR_{\, a} \rr^{ 2}, 
\quad 
\cM_{\, 1} \, \cM_{\, 2} \, \cM_{\, 3} \, \cM_{\, 4} =  \ll \cR_{\, a} \rr^{ 2} \, e^{\, - 4 \, \eta \, \cH}, 
\end{equation}

\noindent so that, 

\begin{equation}
P \ll a, b, b \pm 1 \, \vert \, p \rr =
\frac{
{\it Trace}
\ll \cS_a \, \ll \cR_{\, a} \rr^{\, 2} \, e^{\, -4 \, \eta \, \cH} \rr}{     
{\it Trace}
\ll          \ll \cR_{\, a} \rr^{\, 2} \, e^{\, -4 \, \eta \, \cH} \rr
}, 
\end{equation}

\noindent and the computation of the complicated $P \ll a, b, b \pm 1 \, \vert \, p \rr$ 
is reduced to that of the much simpler $\cH$. 

\subsubsection{Step 3. Show that the elements of $\cH$ are integer multiples of $2 \pi / \tau$}
Setting $u = \eta$ in the first equation in (\ref{simplification.01}), and 
using the first equation in (\ref{simplified.corner.transfer.matrices}), 
it is clear that, 

\begin{equation}
\cQ_{\, 1} \, \cM_{\, 1} = \cQ_{\, 2} \, e^{- \eta \, \cH}, 
\end{equation}

\noindent from which it follows that,  

\begin{equation}
\cA \ll u \rr = \cQ_{\, 2} \, e^{\ll u - \eta \rr \cH} \, \cQ_{\, 2}^{-1}
\label{diagonalized.corner.transfer.matrix}
\end{equation}

\noindent Since $\cA \ll u \rr$ is a function of elliptic theta functions of nome $p = e^{\, - 2 \, \pi / \tau}$, 
the exponential $e^{\ll u - \eta \rr \cH}$ can be calculated using the quasi-periodicity properties of $\cA \ll u \rr$, 
and the elements of $\cH$ must be integer multiples of $2 \, \pi/ \tau$. These integers can be computed by taking 
a suitable limit.

\subsubsection{Remark} 
\label{exponential}
The fact that (the essential factor, that carries the statistical mechanical 
information) in the corner transfer matrix is an exponential, as in 
(\ref{diagonalized.corner.transfer.matrix}), is arguably the most important property of the corner 
transfer matrix, and the one that underlies the result in this work. 

\subsubsection{Step 4. Perform a conjugate modulus transformation}
To take the limit that allows one to compute the elements of $\cH$, Andrews \textit{et al.} perform a conjugate 
modulus transformation that replaces the nome $p$ of the theta functions that appear in the weights with a conjugate 
nome $q$ such that, 

\begin{equation}
p = e^{\, - \, 2 \, \pi / \tau},  
\quad
q = e^{\, - \, 4 \, \pi \, \tau / \ll m + 1 \rr}, 
\quad
\tau \in \RR
\label{modular.parameter.conjugation}
\end{equation}

\subsubsection{Step 5. Compute $\cH$ in unevaluated form}
Working in terms of elliptic theta functions with conjugate nome $q$, Andrews \textit{et al.} found 
that $e^{\, - \, 4 \, \eta \, \cH}$ takes the form, 

\begin{equation}
\ll e^{\, - \, 4 \, \eta \, \cH} \, \rr_{\h, \, \h^{\, \prime}} = 
q^{\, \ll \Phi \ll \h \rr - (2a - m - 1)^2 / 16 \, (m + 1) \rr} \, 
\delta \ll \h, \, \h^{\, \prime} \rr,
\end{equation}

\noindent where $\Phi \ll \h \rr$ is the $L \rightarrow \infty$ limit of the corner 
transfer matrix energy function $\Phi_{\, L} \ll h_0, \cdots, h_{L+1} \rr$, 

\begin{equation}
\Phi_{\, L} \ll h_0 , \cdots, h_{L+1} \rr = 
\sum_{j=1}^L \, j \, \H \ll h_{j-1}, h_{j}, h_{j+1} \rr, 
\quad 
\H \ll h_{j-1}, h_{j}, h_{j+1} \rr = \frac14 \, \vert \, h_{j-1} - h_{j+1}\, \vert
\label{corner.transfer.matrix.energy.function}
\end{equation}

\noindent Writing $\ll \cR_{\, a} \rr_{\h, \, \h^{\, \prime}}$ in terms of $q$, 

\begin{multline}
\ll \cR_{\, a}                          \rr_{\h, \, \h^{\, \prime}} = 
\tau \, q^{\,\ll h_0  - m -1 \rr^2 / 8 \ll m + 1 \rr} E \ll q^{\, a / 2}, q \rr \, 
\delta \ll \h, \, \h^{\, \prime}  \rr,
\\ 
E \ll z, q \rr = \prod_{n = 1}^{\infty} \ll 1 - q^{\, n-1} z          \rr
                                        \ll 1 - q^{\, n-1} z^{\, - 1} \rr
			                            \ll 1 - q^{\, n  }            \rr,  
\label{E}
\end{multline}

\noindent Andrews \textit{et al.} obtain the local height probability in terms of $q$, 

\begin{equation}
P \ll a, b, b \pm 1 \, \vert \, q \rr =
\frac{
E \ll q^{\, a/2}, q^{\, \ll m + 1 \rr /2} \rr \, X \ll a, b, b \pm 1 \, \vert \, q \rr
}{
\sum_{1 \, \leq \, a^{\, \prime} \, < \, m}
E \ll q^{\, a^{\, \prime}/2}, q^{\, \ll m + 1 \rr /2}   \rr \, 
X \ll a^{\, \prime}, b, b \pm 1 \, \vert \, q \, \rr
}, 
\label{P}
\end{equation}

\noindent where $X$ is a one-dimensional configuration sum defined as 
\footnote{\,
We choose to make the dependence on $q$ explicit, as this will change in the sequel.}, 

\begin{equation}
X   \ll a, b, b \pm 1  \, \vert \, q \rr = 
\lim_{\, L_{\, \pi_{ab}} \, \rightarrow \, \infty} X_{\, L} \ll a, b, b \pm 1 \, \vert \, q \rr, 
\label{limit.of.X_L}
\end{equation}

\noindent $\pi_{ab}$ is the parity of $|\, a - b \, |$, and the finite-length one-dimensional 
sum $X_{\, L} \ll a, b, b \pm 1 \, \vert \, q\rr$ is,  

\begin{equation}
X_{\, L} \ll a, b, b \pm 1 \, \vert \, q\rr = 
\sum_{
\ll h_{\, 1}, \cdots, \, h_{L-1} \rr \in 
\ll   1, \cdots, \, m       \rr
} 
q^{\, \Phi_{\, n} \ll h_0 , \cdots, h_{L+1} \rr}, 
\quad 
h_i - h_{i+1} = \pm 1, 
\label{path.partition.function}
\end{equation}

\noindent with the boundary state-variables fixed at $h_0  = a, h_{L} = b, h_{L+1} = b \pm 1$.
The limit in (\ref{limit.of.X_L}) is taken over all even values $L_{\, even}$, or odd values 
$L_{\, odd}$ of $L$, depending on the parity $\pi_{ab}$ of $|\, a - b \, |$. 
The one-dimensional configuration sum $X \ll a, b, b \pm 1\, \vert \, q\rr$, encodes the statistical 
mechanics of the model.

\subsubsection{Remark} Aside from a boundary factor that depends only on the state-variable $a$ 
at the center of the lattice, all information about the statistical mechanics of 
$P \ll a, b, b \pm 1 \, \vert \, q \rr$ is encoded in the one-dimensional configuration sum 
$X \ll a, b, b \pm 1 \, \vert \, q \rr$, which is a $q$-series in the conjugate nome $q$, that 
encodes the multiplicities and energy-levels of the states that contribute to  
$P \ll a, b, b \pm 1 \, \vert \, q \rr$. 

\subsubsection{Step 6. Evaluate the one-dimensional configuration sums} Andrews \textit{et al.}
evaluate the one-dimensional configuration sums $X \ll a, b, b \pm 1\, \vert \, q\rr$ by solving 
the recurrence relations that determine them completely, subject to the initial conditions that 
they satisfy. They obtain the solution in the form of an alternating-sign $q$-series
\footnote{\,
These alternating $q$-series are also known as the Rocha-Caridi Virasoro characters 
\cite{rocha.caridi}
},

\begin{multline}
X   \ll a, \, b, \, b \, \pm \, 1  \, \vert \, q \rr = 
\frac{q^{\, \frac14 \, b \ll b \, \pm \, 1 \rr}}{\eta \ll q \rr} \, 
\Delta \ll a, \frac12 \ll b + (b \pm 1) - 1 \rr \, | \, q \rr, 
\quad 
\eta \ll q \rr = \Pi_{\, n \, = \, 1}^{\, n \, = \, \infinity} \ll 1 - q^{\, n} \rr, 
\\ 
\Delta \ll \, \alpha, \, \beta \, | \, q \, \rr = 
\sum_{n \in \ZZ} 
q^{\, m \ll m + 1 \rr n^2 + \beta \, \ll m+1 \rr \, n + \frac14 \, \alpha \, \ll \alpha - 1 \rr}
\, 
\ll
q^{\, - \, \alpha \, m \, n \, - \, \frac12 \, \alpha \, \beta} - 
q^{     \, \alpha \, m \, n \, + \, \frac12 \, \alpha \, \beta}
\rr
\label{evaluated.sum}
\end{multline}

\subsection{Modular functions} 
\subsubsection{The one-dimensional configurations sums are modular functions}
The $q$-series 
$\Delta \ll \, \alpha, \, \beta \, | \, q \, \rr$ in (\ref{evaluated.sum}) is essentially the difference of two theta functions, 
and the result of the evaluation is that the one-dimensional configuration sums are modular functions. In the case of the hard-hexagon 
model, $q$-series identities due to Schur relate these modular functions to $q$-series with non-negative coefficients that appear as 
sum-sides of Rogers-Ramanujan identities \cite{baxter.1981, andrews.1981}
\footnote{\,
This observation led Andrews, Baxter and Forrester to develop the restricted solid-on-solid models 
as extensions of the hard-square version of the hard-hexagon model, and also led Date, Jimbo, Miwa 
and Okado to their work on the connection between off-critical one-dimensional sums and Virasoro 
characters as discussed in section \textbf{\ref{section.affine.virasoro.characters}}
}.

\subsubsection{Remark. Paths}

\begin{figure}
\begin{tikzpicture}[scale=1.0]

\draw [thin] (0.00, 3.0)--(08.00, 3.0);
\draw [thin] (0.00, 2.0)--(08.00, 2.0);
\draw [thin] (0.00, 1.0)--(08.00, 1.0);
\draw [thin] (0.00, 0.0)--(08.00, 0.0);

\draw [thin] ( 0.0, 0.0)--( 0.0, 3.0);
\draw [thin] ( 1.0, 0.0)--( 1.0, 3.0);
\draw [thin] ( 2.0, 0.0)--( 2.0, 3.0);
\draw [thin] ( 3.0, 0.0)--( 3.0, 3.0);
\draw [thin] ( 4.0, 0.0)--( 4.0, 3.0);
\draw [thin] ( 5.0, 0.0)--( 5.0, 3.0);
\draw [thin] ( 6.0, 0.0)--( 6.0, 3.0);
\draw [thin] ( 7.0, 0.0)--( 7.0, 3.0);
\draw [thin] ( 8.0, 0.0)--( 8.0, 3.0);

\draw [thick] (0.0, 0.0)--(1.0, 1.0);
\draw [thick] (1.0, 1.0)--(2.0, 2.0);
\draw [thick] (2.0, 2.0)--(3.0, 1.0);
\draw [thick] (3.0, 1.0)--(4.0, 2.0);
\draw [thick] (4.0, 2.0)--(5.0, 1.0);
\draw [thick] (5.0, 1.0)--(6.0, 2.0);
\draw [thick] (6.0, 2.0)--(7.0, 1.0);
\draw [thick] (7.0, 1.0)--(8.0, 2.0);

\foreach \x in {0.0,...,8.0}
{
\draw [fill=black!50] (\x,3) circle (0.08);
\draw [fill=black!50] (\x,2) circle (0.08);
\draw [fill=black!50] (\x,1) circle (0.08);
\draw [fill=black!50] (\x,0) circle (0.08);
}

\node at (-0.50, 3.00) {$\textswab{4}$};
\node at (-0.50, 2.00) {$\textswab{3}$};
\node at (-0.50, 1.00) {$\textswab{2}$};
\node at (-0.50, 0.00) {$\textswab{1}$};

\node at (8.50, 2.50) {$\textswab{3}$};
\node at (8.50, 1.50) {$\textswab{2}$};
\node at (8.50, 0.50) {$\textswab{1}$};

\end{tikzpicture}
\caption{\textit{
A minimal path in the restricted solid-on-solid model $\cL_{\, 4, 5}$, and this 
particular path belongs to the one-dimensional configuration sum labeled by $a=1$, 
$b=2$, and $c=3$. 
The numbers on the left label the possible initial points of paths, 
and correspond to the label $s$ in $\chi^{\, Vir}_{\, r, \, s}$.
The numbers on the right label the possible final bands that paths can end up 
oscillating in, 
and correspond to the label $r$ in $\chi^{\, Vir}_{\, r, \, s}$.
}}
\label{figure.path.01}
\end{figure}
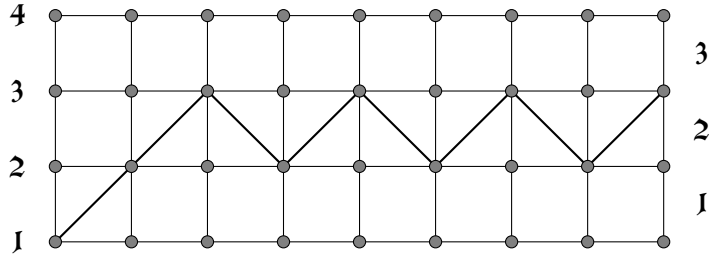

\begin{figure}
\begin{tikzpicture}[scale=1.0]

\draw [thin] (0.00, 3.0)--(08.00, 3.0);
\draw [thin] (0.00, 2.0)--(08.00, 2.0);
\draw [thin] (0.00, 1.0)--(08.00, 1.0);
\draw [thin] (0.00, 0.0)--(08.00, 0.0);

\draw [thin] ( 0.0, 0.0)--( 0.0, 3.0);
\draw [thin] ( 1.0, 0.0)--( 1.0, 3.0);
\draw [thin] ( 2.0, 0.0)--( 2.0, 3.0);
\draw [thin] ( 3.0, 0.0)--( 3.0, 3.0);
\draw [thin] ( 4.0, 0.0)--( 4.0, 3.0);
\draw [thin] ( 5.0, 0.0)--( 5.0, 3.0);
\draw [thin] ( 6.0, 0.0)--( 6.0, 3.0);
\draw [thin] ( 7.0, 0.0)--( 7.0, 3.0);
\draw [thin] ( 8.0, 0.0)--( 8.0, 3.0);

\draw [thick] (0.0, 0.0)--(1.0, 1.0);
\draw [thick] (1.0, 1.0)--(2.0, 0.0);
\draw [thick] (2.0, 0.0)--(3.0, 1.0);
\draw [thick] (3.0, 1.0)--(4.0, 2.0);
\draw [thick] (4.0, 2.0)--(5.0, 3.0);
\draw [thick] (5.0, 3.0)--(6.0, 2.0);
\draw [thick] (6.0, 2.0)--(7.0, 1.0);
\draw [thick] (7.0, 1.0)--(8.0, 2.0);

\foreach \x in {0.0,...,8.0}
{
\draw [fill=black!50] (\x,3) circle (0.08);
\draw [fill=black!50] (\x,2) circle (0.08);
\draw [fill=black!50] (\x,1) circle (0.08);
\draw [fill=black!50] (\x,0) circle (0.08);
}

\node at (-0.50, 3.00) {$\textswab{4}$};
\node at (-0.50, 2.00) {$\textswab{3}$};
\node at (-0.50, 1.00) {$\textswab{2}$};
\node at (-0.50, 0.00) {$\textswab{1}$};

\node at (8.50, 2.50) {$\textswab{3}$};
\node at (8.50, 1.50) {$\textswab{2}$};
\node at (8.50, 0.50) {$\textswab{1}$};

\end{tikzpicture}
\caption{\textit{
A non-minimal path in the restricted solid-on-solid model $\cL_{\, 4, 5}$, that  
belongs to the one-dimensional configuration sum labeled by $a=1$, $b=2$, and $c=3$. 
}}
\label{figure.path.02}
\end{figure}
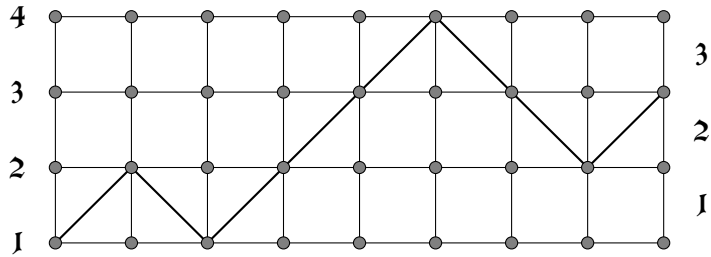

The one-dimensional configuration sums are partition functions of paths on a Dynkin diagram 
that can be represented as paths on lattice as in Figures \textbf{\ref{figure.path.01}}, and
\textbf{\ref{figure.path.02}}. 
The path in Figure \textbf{\ref{figure.path.01}} belongs to a configuration sum labelled by 
$a=1$, $b=2$, and $c=3$. 
It is a minimal path in the sense that it starts from $a=1$, and reaches $b=2$, to oscillate 
in the band $\ll b, c \rr$ = $\ll 2, 3 \rr$, in a minimal number of steps. 
The path in Figure \textbf{\ref{figure.path.02}} belongs to the same configuration sum 
labelled by $a=1$, $b=2$, and $c=3$, as in Figure \textbf{\ref{figure.path.01}}, but is  
non-minimal          in the sense that it starts from $a=1$, and reaches $b=2$, to oscillate 
in the band $\ll b, c \rr$ = $\ll 2, 3 \rr$, in a non-minimal number of steps. 
Computing the energy of a path using (\ref{path.partition.function}), the minimal path in 
a particular one-dimensional configuration sum has a lower energy than any other path. 

\subsubsection{Remark. $q$-Series representations of the one-dimensional configuration sums with 
positive-definite coefficients}
Following the work of Kedem, Klassen, McCoy and Melzer \cite{kedem.klassen.mccoy.melzer.01,
kedem.klassen.mccoy.melzer.02} on the positive-definite representations of the Virasoro 
characters, extensive studies of the path representations of the one-dimensional configuration sums, 
including 
\cite{berkovich, berkovich.mccoy, warnaar.01, warnaar.02, foda.lee.welsh, foda.lee.pugai.welsh, 
foda.welsh, welsh} and references therein, led to evaluations of large classes of Virasoro characters 
in terms of $q$-series with positive-definite coefficients.
 
\section{Off-critical local height probabilities in a planar geometry are affine and 
Virasoro characters}
\label{section.affine.virasoro.characters}
\textit{We recall the definitions of affine and Virasoro characters, then following 
\cite{date.jimbo.miwa.okado.01}, we write the local height probabilities in terms of 
characters.}

\subsection{Affine characters}

\subsubsection{Theta functions with characteristics}
Following \cite{mumford.01}, the characteristic-$i$, level-$j$ theta function is, 

\begin{equation}
\theta_{\, i,   \, j}   \ll z, \, q \rr =  
\sum_{n \, \in \, \ZZ} 
q^{\, j \, \ll n \, + \, \frac{\, i}{2 j} \rr^2} 
z^{\, j \, \ll n \, + \, \frac{\, i}{2 j} \rr}, 
\label{theta.function.with.characteristics}
\end{equation}

\noindent where 
$i \in \ll 0, 1, \cdots, j \rr$, 
$j \in \ll 1, 2, \cdots    \rr$,
$q$ is the nome parameter, and $z \in \CC$. 

\subsubsection{A linear combination of theta functions with characteristics}
Following \cite{kac.peterson}, we define the linear combination of theta 
functions with characteristsics
\footnote{\,
This is the second equation in (7), in \cite{date.jimbo.miwa.okado.01}, with appropriate 
changes in notation.
}, 

\begin{equation}
\Theta_{\, i,   \, j}   \ll z, \, q \rr =  
\sum_{n \, \in \, \ZZ} 
q^{\, j \, \ll n \, + \, \frac{\, i}{2 j} \rr^2} 
\ll z^{\, - \, j \, \ll n \, + \, \frac{\, i}{2 j} \rr} - 
    z^{\,      j \, \ll n \, + \, \frac{\, i}{2 j} \rr} \rr
\label{linear.combination.theta.function.with.characteristics}
\end{equation}

\subsubsection{Affine characters}
Following Kac and Peterson \cite{kac.peterson}, the character of the charge-$i$, level-$j$ 
$\cU_q \ll \widehat{sl_{\, 2}} \rr$ irreducible highest-weight module can be written in terms 
of theta functions as
\footnote{\,
First equation in (7), in \cite{date.jimbo.miwa.okado.01}, with appropriate changes in notation. 
Note the shift in the indices of the theta function in the numerator of the right hand side with 
respect to those of the affine character on the left hand side.
}, 

\begin{equation}
\widehat{\chi}_{\, i, \, j} \ll z, q \rr = 
\frac{ 
\Theta_{\, i + 1, \, j + 2} \ll z, \, q \rr}{
\Theta_{\,     1, \,     2} \ll z, \, q \rr}, 
\label{affine.character}
\end{equation}

\noindent where $\Theta_{\, i + 1, \, j + 2} \ll z, \, q \rr$ is defined in 
(\ref{linear.combination.theta.function.with.characteristics}). 

\subsection{Virasoro characters}

\subsubsection{Virasoro characters as branching functions}
Based on the work of Kac and Peterson \cite{kac.peterson}, on affine Lie algebras and 
theta functions, Date, Jimbo, Miwa, and Okado show, in \cite{date.jimbo.miwa.okado.01}, 
that the product of two affine characters decomposes in the form,  

\begin{equation}
\widehat{\chi}_{\, j_{\, 1}, \, m-2} \ll z, \, q \rr            
\widehat{\chi}_{\, j_{\, 2}, \,   1} \ll z, \, q \rr  
=
\sum_{j_{\, 3}} \, c_{\, j_{\, 1} \, j_{\, 2} \, j_{\, 3}} \ll q \rr 
\widehat{\chi}_{\, j_{\, 3}, \, m-1} \ll \, z, \, q \rr, 
\label{branching.functions}
\end{equation}

\noindent where the branching coefficients 
$c_{\, j_{\, 1} \, j_{\, 2} \, j_{\, 3}} \ll q \rr$ are characters of the fully-degenerate 
irreducible Virasoro highest-weight modules that appear in the unitary rational conformal 
field theories of Friedan, Qiu and Shenker \cite{friedan.qiu.shenker.01, friedan.qiu.shenker.02}
\footnote{\,
Note that the level indices of the affine characters are such that, for $m=3$, we get the 
Virasoro characters of the coset model that corresponds to the Ising model. 
},
in parallel with the coset construction of these Virasoro modules in terms of 
$\cU_q \ll \widehat{sl_{\, 2}} \rr$ highest-weight modules by 
Goddard, Kent and Olive \cite{goddard.kent.olive.01, goddard.kent.olive.02}. 

\subsubsection{Notation}
The notation $c_{\, j_{\, 1} \, j_{\, 2} \, j_{\, 3}} \ll q \rr$ used for the branching coefficients
of products of affine characters in (\ref{branching.functions}), which are Virasoro characters, 
is related to 
the notation $\chi^{\, Vir}_{\, r, \, s} \ll q \rr$ that is currently more commonly used for Virasoro 
characters, as follows. For $c=b+1$, 

\begin{equation}
c_{\, j_{\, 1} \, j_{\, 2} \, j_{\, 3}} \ll q \rr = 
c_{\, b, \, 1, \, a}                    \ll q \rr = 
\chi^{\, Vir}_{\, b, \, a}              \ll q \rr, 
\label{notation.01}
\end{equation}

\noindent and for $c=b-1$,

\begin{equation}
c_{\, j_{\, 1} \, j_{\, 2} \, j_{\, 3}} \ll q \rr = 
c_{\, c, \, 2, \, a}                    \ll q \rr = 
\chi^{\, Vir}_{\, c, \, a}              \ll q \rr 
    \label{notation.02}
\end{equation}

\noindent In other words, 
the first index $j_{\, 1}$ in $c_{\, j_{\, 1} \, j_{\, 2} \, j_{\, 3}}$ is equal to 
the first index $r$        in $\chi^{\, Vir}_{\, r, \, s}$, 
and takes the value of the smaller of the outer boundary state-variables $\ll b, c \rr$, 
the second index $j_{\, 2}$ in $c_{\, j_{\, 1} \, j_{\, 2} \, j_{\, 3}}$ is either 1 or 2 
depending on whether $c=b+1$ or $b-1$, respectively, and 
the third  index $j_{\, 3}$ in $c_{\, j_{\, 1} \, j_{\, 2} \, j_{\, 3}}$ is equal to 
the second index $s$ in $\chi^{\, Vir}_{\, r, \, s}$, and takes the value of the inner 
boundary state-variable $a$. 
 
\subsubsection{One-dimensional configuration sums are Virasoro characters}
\label{work.date.jimbo.miwa.okado}
In \cite{date.jimbo.miwa.okado.01, date.jimbo.miwa.okado.02}, 
Date, Jimbo, Miwa and Okado observe that the regime-III one-dimensional sum 
$X \ll a, b, b \pm 1 \, \vert \, q \rr$ is identical to the character of a Virasoro 
highest-weight module in a unitary minimal conformal field theory, up to a $q$-factor
\footnote{\,
Further details with full proofs can be in the work of Date \textit{et al.} with Kuniba 
\cite{date.jimbo.kuniba.miwa.okado.03, date.jimbo.kuniba.miwa.okado.04}, 
}, 
 
\begin{equation}
X \ll a, b, b \pm 1 \, \vert \,  q \rr = 
q^{\, \nu} \, c_{\, j_{\, 1}\, j_{\, 2}\, j_{\, 3}} \ll q \rr,  
\end{equation}

\noindent where, 

\begin{multline}
\nu =
\frac14 
\ll 
\ll b-a \rr 
- 
\frac{ j_{\, 1}^{\, 2} }{m}
- 
\frac{ j_{\, 2}^{\, 2} }{3}
+ 
\frac{ j_{\, 3}^{\, 2} }{\ll m + 1 \rr}
+ 
\frac12
\rr, 
\\
j_{\, 1} = \ll \frac{b+c}{2} \rr - \frac12, \quad 
j_{\, 2} = \ll \frac{b-c}{2} \rr + \frac32, \quad 
j_{\, 3} = a
\end{multline}

\noindent The branching coefficients $c_{\, j_, \, j_2 \, j_3}$ in (\ref{branching.functions}) are 
characters of the Virasoro highest-weight modules that appear in the unitary minimal conformal field 
theories $\cM_{\, m, \, m+1}$ of 
Friedan, Qiu and Shenker \cite{friedan.qiu.shenker.01, friedan.qiu.shenker.02}.
$\cM_{\, m, \, m+1}$ were identified by Huse \cite{huse} as the {\it critical} limit of the 
$\cL_{\, m, \, m+1}$ solid-on-solid models \cite{andrews.baxter.forrester}, that we are interested 
in in this work. 
The affine characters $\widehat{\chi}_{\, i, j}$ in (\ref{branching.functions}) are characters of 
the $\widehat{sl_{\, 2}}$ Wess-Zumino-Witten models used by Goddard, Kent and Olive 
\cite{goddard.kent.olive.01, goddard.kent.olive.02} to construct $\cM_{\, m, \, m+1}$.  

\subsection{Re-writing the local height probabilities in terms of affine and Virasoro characters}

Following \cite{date.jimbo.miwa.okado.01}, we use the identity
\footnote{\,
Equation (2.13) in \cite{date.jimbo.kuniba.miwa.okado.03}, with 
$x^2 \rightarrow q$, 
$j   \rightarrow i$,
$m   \rightarrow j$, 
$\epsilon_1 = -$, 
$\epsilon_2 = +$,
dropping the superscripts off $\Theta$ in (2.13), and re-arranging terms.
}, 

\begin{equation}
E \ll q^{i/2}, q^{j/2} \rr = 
q^{\frac{i \ll j - i \rr}{4 j}} \, \Theta_{\, i, \, j} \ll q^{\, \frac12}, \, q \rr,   
\end{equation}

\noindent as well as (\ref{affine.character}--\ref{branching.functions}), to rewrite (\ref{P}) 
in terms of theta functions and Virasoro characters,  

\begin{equation}
P \ll a, b, b \pm 1 \, \vert \, q \rr = 
\ll 
q^{- \nu}
\frac{
\Theta_{\, j_3 + 1, \, m + 1} \ll q^{1/2}, \, q \rr \Theta_{\,   1, \, 2} \ll q^{1/2}, \, q \rr }{
\Theta_{\, j_1 + 1, \,     3} \ll q^{1/2}, \, q \rr \Theta_{\, j_2 + 1, \, m} \ll q^{1/2}, \, q \rr  
}
\rr 
\ll 
q^{\nu} c_{\, j_{\, 1}\, j_{\, 2}\, j_{\, 3}} \ll q \rr 
\rr, 
\label{rewriting}
\end{equation}

\noindent where $\ll j_{\, 1}, \, j_{\, 2}, \, j_{\, 3} \rr$ are related to 
$\ll a, \, b, \, c = b \pm 1 \rr$ as in (\ref{notation.01}--\ref{notation.02}), and 
the $q^{\, \pm \nu}$-factors are positioned to make the origin of the various factors 
clear.    

\section{Critical partition functions in a cylindrical geometry}
\label{section.other.works}
\textit{We recall basic works related to our understanding of the critical partition functions in 
a cylindrical geometry as Virasoro characters.}

\subsection{Numerical results}
\label{work.gehlen.rittenberg}
In \cite{gehlen.rittenberg}, Gehlen and Rittenberg study the spectra of the quantum Hamiltonians of the Ising (2-state Potts) 
and the 3-state Potts spin-chains on a finite-height, infinite-circumference cylindric-geometry, with free boundary conditions 
in the finite-height direction
\footnote{\,
In \cite{gehlen.rittenberg}, this geometry is referred to as a strip geometry with a finite-direction and an infinite-direction, 
and periodic boundary conditions in the latter direction.
}.
In the Ising case, they use the exact diagonalization of the Ising quantum Hamiltonian of \cite{boccara.sarma, burkhardt.guim}, 
and in the 3-state Potts case, they diagonalize the quantum Hamiltonian numerically. 
In both cases, they show that the scaling operators fall into fully-degenerate irreducible Virasoro highest-weight modules of 
a single Virasoro algebra, as opposed to the direct product of two commuting Virasoro algebras as in the same models on a domain 
with periodic boundary conditions \cite{cardy.01}, and the corresponding partition functions of these models with these boundary 
conditions are linear (as opposed to bilinear) sums of characters of these modules
\footnote{\,
That one Virasoro algebra (as opposed to two commuting Virasoro algebras) is needed in the presence of a boundary was expected 
on the basis of analogous computations in open (as opposed to closed) string theories \cite{superstrings.book}
}. 

\subsection{Analytic results}
\label{work.cardy}
In \cite{cardy.02}, Cardy used the inversion sum rules, that follow from requiring consistency under modular transformations, to 
study the spectra of the transfer matrices of the unitary minimal Virasoro conformal field theories of Friedan, Qiu and Shenker
\cite{friedan.qiu.shenker.01, friedan.qiu.shenker.02}, in the same finite-height, infinite-circumference cylindric-geometry of 
\cite{gehlen.rittenberg}, with 
\1 free boundary conditions as in \cite{gehlen.rittenberg}, 
\2 fixed boundary conditions (the order parameter is fixed to the same value on both boundaries, or two different values on the 
two boundaries), and 
\3 mixed boundary conditions (fixed boundary conditions on one boundary and free boundary conditions on the other), 
in the finite-direction. 
Cardy obtained complete analytic results for the partition functions of these models, with these boundary conditions, as linear 
sums of the full-degenerate irreducible Virasoro highest-weight modules of a single Virasoro character, in agreement with, and 
extension of \cite{gehlen.rittenberg}.

\subsection{Solid-on-solid and restricted-solid-on-solid results}
\label{saleur.bauer.work}
In \cite{saleur}, Saleur computed the partition function of the solid-on-solid version of 
the six-vertex model on a cylinder with 
the bottom-row of state-variables fixed at $a \in \ZZ$, and
the top-row    of state-variables fixed at $c \in \ZZ$, for admissible choices of $a$ and $c$. 
In \cite{saleur.bauer}, Saleur and Bauer extended the results in \cite{saleur} to two classes of models
on a cylinder of dimensions $H \times W$, where $W$ is the circumference of the periodic direction, with more general boundary 
conditions in the non-periodic direction of length $H$, namely 
\1 the critical solid-on-solid version of the six-vertex model on a cylinder with boundary conditions 
such that
the bottom-row          of state-variables is fixed at height $a \in \ZZ$,    
the second-from-top-row of state-variables is fixed at height $b \in \ZZ$, and  
the top-row             of state-variables is fixed at height $c \in \ZZ$, $b - c = \pm 1$, with admissible 
choices of $a, b$ and $c$, given the number of rows $H$ in the non-periodic direction of the cylinder, 
and
\2 the critical version of the restricted solid-on-solid models of Andrews, Baxter and Forrester
\cite{andrews.baxter.forrester}, and the critical solid-on-solid models of Pasquier \cite{pasquier.01}, 
on a cylinder with boundary conditions as in \textbf{1}, but with the state-variables $a, b$ and $c$ suitably 
restricted.

\subsubsection{The Bethe Ansatz} 
Because the boundary conditions in \cite{saleur.bauer} are more elaborate than those in \cite{saleur}, Saleur and Bauer use, 
not only the Coulomb gas representation of restricted solid-on-solid models \cite{pasquier.02}, but also the Bethe Ansatz analysis
of \cite{hamer.quispel.batchelor, alcaraz.barber.batchelor}. They find that the critical partition functions of the models 
that they consider on a cylinder are Virasoro characters labelled by $a, b$ and $c = b \pm 1$. 

\subsubsection{Relation to the work of Date \textit{et al.}}
Saleur and Bauer find that the spectrum of the \textit{off-critical} regime-III corner transfer matrix Hamiltonian is, 
up to a shift in the ground state eigenvalue, the same as the spectrum of the corresponding \textit{critical} continuum-limit 
Hamiltonian on a cylinder, for the same fixed boundary conditions, \textit{if} the deviation from criticality in the former model 
is identified with the finite-size $L$ of the non-periodic direction of the cylinder in the latter 
\footnote{\,
In \cite{saleur.bauer}, Saleur and Bauer also study critical and tri-critical Potts model partition functions 
on a cylinder and 
obtain results in the form of linear combinations of Virasoro characters in agreement with, and extension of Gehlen and Rittenberg's 
numerical results \cite{gehlen.rittenberg}, and Cardy's analytic results \cite{cardy.02}. Please note that reference 
[14] in \cite{saleur.bauer} is incorrect, and should be replaced with reference \cite{cardy.02} in the present work. 
}.
Recalling the result of Date \textit{et al.} \cite{date.jimbo.miwa.okado.01, date.jimbo.miwa.okado.02}, 
Saleur and Bauer found concrete examples to the effect that off-critical local height probabilities, 
in type-$A$ and type-$D$ regime-III restricted solid-on-solid models on a plane, are essentially the 
same objects as the partition functions of critical conformal field theories on a cylinder, for 
matching choices of boundary state-variables $a, b$ and $c$, and a suitable mapping of the the role of departure 
from criticality in the off-critical problem on a plane to the finite-size in the critical problem
on a cylinder.

\smallskip

\textit{The present work motivates why the aspect-ratio of the annular, or equivalently cylinder  
geometry of the critical conformal field theories can be identified with the distance from criticality 
in the corresponding off-critical statistical mechanical models.}  

\subsection{Restricted-solid-on-solid models on a spiral geometry} 
\label{work.present.work}
The main idea in the present work is 
\1 to extend the definition of the local height probability 
$P \ll a, b, b \pm 1 \, \vert \, p \rr$ from the 4-quadrant planar geometry to a $4N$-quadrant 
spiral geometry, where $N = \frac54, \frac64, \cdots$, and at the same time
\2 allow the modular parameter $\tau$, whose inverse $\tau^{\, \prime} = \frac{1}{\tau}$ measures 
departure from criticality, to depend on $N$, and write the new $N$-dependent modular parameter 
as $\tau_N$. This way, we obtain an object that interpolates the off-critical local height 
probabilities on a plane and the critical partition functions on a cylinder. 

\section{Evaluating local height probabilities in a spiral geometry}
\label{section.spiral.geometry}
\textit{We repeat the derivation in section \textbf{\ref{evaluating.the.local.height.probabilities}} 
of the off-critical local height probabilities, but now in a spiral geometry. 
We show that all steps carry through with minimal modification. }

\subsection{From an off-critical system on a plane to a near-critical system on 
a \lq deep\rq\, spiral} 
\label{step.01}
We show that the computation of the local height probabilities of Andrews \textit{et al.} extends 
to a $4N$-quadrant geometry without obstruction. Starting from an off-critical system with weights 
in terms of elliptic theta functions, with a modular parameter $\tau$, and increasing the number 
of quadrants to $4N$, then aside from factors that do not carry information about the statistical 
mechanics of the local height probabilities, we essentially only change the modular parameter from 
$\tau$ to $\tau_{\, N} = \tau / N $. 
In other words, we find that the essential factor in the local height probability, evaluated
on a $4 N$-quadrant geometry, which is the one-dimensional configuration sum, becomes 
$X \ll a, b, b \pm 1 \, \vert \, q^{\, N} \rr$, where  
$q = e^{\, - \, 4 \, \pi \, \tau / \ll m + 1 \rr}$. 
Now if we also vary $\tau$, make that $\tau_N$, and choose $\tau_N = \tau / N$, then the 
one-dimensional configuration sums remain invariant. 
This means that, apart from overall factors that are blind to the statistical mechanics, 
the off-critical local height probability on a plane is equal to that on a spiral with 
$4N$-quadrants. 
For sufficiently-large but finite $N$, the spiral is \textit{\lq deep\rq}, in the sense that 
it has many quadrants, that is a large number of turns, but still finite, and the model 
that lives on it is near-criticality, in the sense that the nome $p$ of the elliptic theta 
functions in which the Boltzmann weights are expressed, is finite but small.

\subsection{$4N$-quadrant geometry}

Following \cite{baxter.book}, Chapter \textbf{13}, we consider a finite version of a square 
lattice as in Figure \textbf{ 1}. We take $N$ copies of the above finite lattice, labelled by 
$i \in \ll 1, 2, \cdots, N \rr$ and {\it cut} each copy along a half-line that extends from 
the center of the lattice to a corner, {\it e.g.} the East corner. This splits each site on 
the cut half-line into two. We denote the upper and lower split half-lines by $h_{i+}$ and 
$h_{i-}$, respectively. We identify $h_{i+}$ and $h_{\ll i+1 \rr -}$ for $i \in $ $\ll 1, 2, 
\cdots, N-1\rr$ and identify $h_{i+}$ to $h_{\ll i+1 \rr -}$ to obtain a $4N$-quadrant spiral 
geometry with periodic boundary conditions in the angular direction.

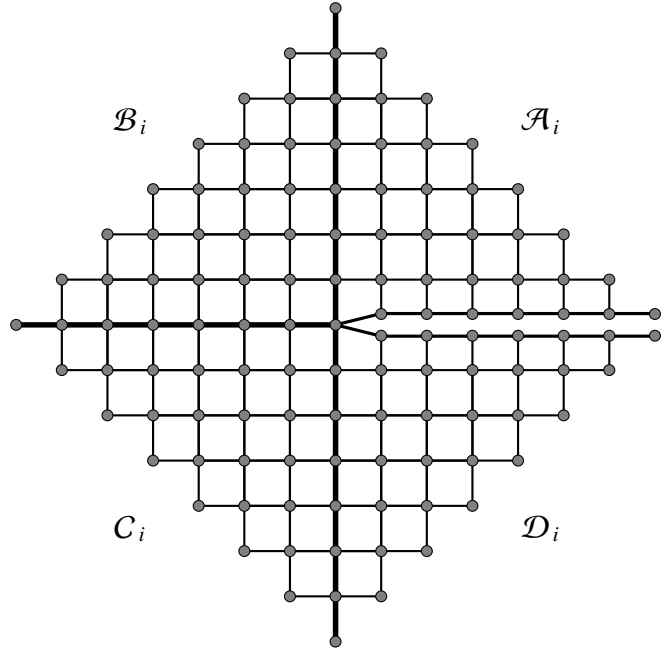
\begin{figure}
\begin{tikzpicture}[scale=.6]


\draw [thick] (0, 5) rectangle (1,6);

\draw [thick] (0, 4) rectangle (1,5);
\draw [thick] (1, 4) rectangle (2,5);

\draw [thick] (0, 3) rectangle (1,4);
\draw [thick] (1, 3) rectangle (2,4);
\draw [thick] (2, 3) rectangle (3,4);

\draw [thick] (0, 2) rectangle (1,3);
\draw [thick] (1, 2) rectangle (2,3);
\draw [thick] (2, 2) rectangle (3,3);
\draw [thick] (3, 2) rectangle (4,3);

\draw [thick] (0, 1) rectangle (1,2);
\draw [thick] (1, 1) rectangle (2,2);
\draw [thick] (2, 1) rectangle (3,2);
\draw [thick] (3, 1) rectangle (4,2);
\draw [thick] (4, 1) rectangle (5,2);

\draw [very thick] (-0.025,-7.0)--(-0.025,7.0);
\draw [very thick] ( 0.025,-7.0)--( 0.025,7.0);

\draw [very thick] (-7.0, -0.025)--(0.0, -0.025);
\draw [very thick] (-7.0,  0.025)--(0.0,  0.025);

\draw  [very thick] (0, 0)--(1,  0.25); 
\draw  [very thick] (0, 0)--(1, -0.25);

\draw  [very thick] (1,  0.25)--(7,  0.25);
\draw  [very thick] (1, -0.25)--(7, -0.25);

\draw  [thick] (5,  1)--(6,  1);
\draw  [thick] (5, -1)--(6, -1);

\draw  [thick] (1,  1)--(1, 0.25); 
\draw  [thick] (2,  1)--(2, 0.25); 
\draw  [thick] (3,  1)--(3, 0.25); 
\draw  [thick] (4,  1)--(4, 0.25); 
\draw  [thick] (5,  1)--(5, 0.25); 
\draw  [thick] (6,  1)--(6, 0.25); 

\draw  [thick] (1,  -1)--(1, -0.25); 
\draw  [thick] (2,  -1)--(2, -0.25); 
\draw  [thick] (3,  -1)--(3, -0.25); 
\draw  [thick] (4,  -1)--(4, -0.25); 
\draw  [thick] (5,  -1)--(5, -0.25); 
\draw  [thick] (6,  -1)--(6, -0.25);

\draw [thick] (0, -2) rectangle (1,-1);
\draw [thick] (1, -2) rectangle (2,-1);
\draw [thick] (2, -2) rectangle (3,-1);
\draw [thick] (3, -2) rectangle (4,-1);
\draw [thick] (4, -2) rectangle (5,-1);

\draw [thick] (0, -3) rectangle (1,-2);
\draw [thick] (1, -3) rectangle (2,-2);
\draw [thick] (2, -3) rectangle (3,-2);
\draw [thick] (3, -3) rectangle (4,-2);

\draw [thick] (0, -4) rectangle (1,-3);
\draw [thick] (1, -4) rectangle (2,-3);
\draw [thick] (2, -4) rectangle (3,-3);

\draw [thick] (0, -5) rectangle (1,-4);
\draw [thick] (1, -5) rectangle (2,-4);

\draw [thick] (0, -6) rectangle (1,-5);


\draw [thick] ( 0, 5) rectangle (-1,6);

\draw [thick] ( 0, 4) rectangle (-1,5);
\draw [thick] (-1, 4) rectangle (-2,5);

\draw [thick] ( 0, 3) rectangle (-1,4);
\draw [thick] (-1, 3) rectangle (-2,4);
\draw [thick] (-2, 3) rectangle (-3,4);

\draw [thick] ( 0, 2) rectangle (-1,3);
\draw [thick] (-1, 2) rectangle (-2,3);
\draw [thick] (-2, 2) rectangle (-3,3);
\draw [thick] (-3, 2) rectangle (-4,3);

\draw [thick] ( 0, 1) rectangle (-1,2);
\draw [thick] (-1, 1) rectangle (-2,2);
\draw [thick] (-2, 1) rectangle (-3,2);
\draw [thick] (-3, 1) rectangle (-4,2);
\draw [thick] (-4, 1) rectangle (-5,2);

\draw [thick] ( 0, 0) rectangle (-1,1);
\draw [thick] (-1, 0) rectangle (-2,1);
\draw [thick] (-2, 0) rectangle (-3,1);
\draw [thick] (-3, 0) rectangle (-4,1);
\draw [thick] (-4, 0) rectangle (-5,1);
\draw [thick] (-5, 0) rectangle (-6,1);

\draw [thick] ( 0, -1) rectangle (-1,0);
\draw [thick] (-1, -1) rectangle (-2,0);
\draw [thick] (-2, -1) rectangle (-3,0);
\draw [thick] (-3, -1) rectangle (-4,0);
\draw [thick] (-4, -1) rectangle (-5,0);
\draw [thick] (-5, -1) rectangle (-6,0);

\draw [thick] ( 0, -2) rectangle (-1,-1);
\draw [thick] (-1, -2) rectangle (-2,-1);
\draw [thick] (-2, -2) rectangle (-3,-1);
\draw [thick] (-3, -2) rectangle (-4,-1);
\draw [thick] (-4, -2) rectangle (-5,-1);

\draw [thick] ( 0, -3) rectangle (-1,-2);
\draw [thick] (-1, -3) rectangle (-2,-2);
\draw [thick] (-2, -3) rectangle (-3,-2);
\draw [thick] (-3, -3) rectangle (-4,-2);

\draw [thick] ( 0, -4) rectangle (-1,-3);
\draw [thick] (-1, -4) rectangle (-2,-3);
\draw [thick] (-2, -4) rectangle (-3,-3);

\draw [thick] ( 0, -5) rectangle (-1,-4);
\draw [thick] (-1, -5) rectangle (-2,-4);

\draw [thick] ( 0, -6) rectangle (-1,-5);

\foreach \x in {0,...,12}
{
\draw [fill=black!50] (\x-6, 1) circle (0.12);
\draw [fill=black!50] (\x-6,-1) circle (0.12);
}

\foreach \x in {0,...,7}
{
\draw [fill=black!50] (\x-7, 0) circle (0.12);
}

\foreach \x in {7,...,13}
{
\draw [fill=black!50] (\x-6, 0.24) circle (0.12);
\draw [fill=black!50] (\x-6,-0.24) circle (0.12);
}

\foreach \x in {0,...,10}
{
\draw [fill=black!50] (\x-5, 2) circle (0.12);
\draw [fill=black!50] (\x-5,-2) circle (0.12);
}

\foreach \x in {0,...,8}
{
\draw [fill=black!50] (\x-4, 3) circle (0.12);
\draw [fill=black!50] (\x-4,-3) circle (0.12);
}

\foreach \x in {0,...,6}
{
\draw [fill=black!50] (\x-3, 4) circle (0.12);
\draw [fill=black!50] (\x-3,-4) circle (0.12);
}

\foreach \x in {0,...,4}
{
\draw [fill=black!50] (\x-2, 5) circle (0.12);
\draw [fill=black!50] (\x-2,-5) circle (0.12);
}

\foreach \x in {0,...,2}
{
\draw [fill=black!50] (\x-1, 6) circle (0.12);
\draw [fill=black!50] (\x-1,-6) circle (0.12);
}

\draw [fill=black!50] (0, 7) circle (0.12);
\draw [fill=black!50] (0,-7) circle (0.12);

\node at ( 4.5, 4.5) {$\cA_{\, i}$};
\node at (-4.5, 4.5) {$\cB_{\, i}$};
\node at (-4.5,-4.5) {$\cC_{\, i}$};
\node at ( 4.5,-4.5) {$\cD_{\, i}$};

\end{tikzpicture}
\caption{
\textit{We cut the 4-quadrant lattice along a half-line, then prepare $N$ copies 
of the cut lattices, labeled by $i \in 1, 2, \cdots, N$. 
}
}
\label{}
\end{figure}

\subsection{Local height probabilities on $4N$-quadrants}

Our task is identical to that in section \textbf{\ref{evaluating.the.local.height.probabilities}}, 
on a $4N$-quadrant lattice, $N \in \NN > 1$, rather than $N=1$, so that now we have $N$ copies of 
each quadrant, $\cA_i (v)$, {\it etc.} 

\subsubsection{Step 1 on a spiral} Using the Yang-Baxter equations, we obtain,

\begin{multline}
\cA_{\, i} \ll u\rr = \cQ_{\, 1, \, i} \, \cM_{\, 1, \, i} \, e^{ \, u \, \cH} \, \cQ_{\, 2, \, i}^{-1},  \quad 
\cB_{\, i} \ll u\rr = \cQ_{\, 2, \, i} \, \cM_{\, 2, \, i} \, e^{-\, u \, \cH} \, \cQ_{\, 3, \, i}^{-1}, \\ 
\cC_{\, i} \ll u\rr = \cQ_{\, 3, \, i} \, \cM_{\, 3, \, i} \, e^{ \, u \, \cH} \, \cQ_{\, 4, \, i}^{-1},  \quad 
\cD_{\, i} \ll u\rr = \cQ_{\, 4, \, i} \, \cM_{\, 4, \, i} \, e^{-\, u \, \cH} \, \cQ_{\, 1, \, i}^{-1} 
\label{simplification.01.4N}
\end{multline}

\noindent where $\cH$ is the same, and the matrices $\cM_i$ and $\cQ_i$, $i = 1, \cdots, N$, have
the same properties as on a plane. This allows us to write, 

\begin{equation}
P_N \ll a, b, b \pm 1 \, \vert \, p_N \rr = 
\frac{
\textit{Trace}  \ll S_a \, M_{1, 1} \, M_{2, 1} \, M_{3, 1} \cdots M_{3, N} \, M_{4, N} \, M_{1, N} \rr}{
\textit{Trace}  \ll        M_{1, 1} \, M_{2, 1} \, M_{3, 1} \cdots M_{3, N} \, M_{4, N} \, M_{1, N} \rr}, 
\end{equation}

\noindent where $P_N \ll a, b, b \pm 1 \, \vert \, p_N \rr$ is the local height probability on a spiral 
geometry with $4N$ quadrants, that is winding number $N$, the notation $p_N$ indicates that we allow the 
nome $p$ to vary with $N$, and we take 
$P_{\, 1} \ll a, b, b \pm 1 \, \vert \, p_1 \rr = P \ll a, b, b \pm 1 \, \vert \, p \rr$. 

\subsubsection{Step 2 on a spiral} The analysis of each separate quadrant is independent 
of the total number of quadrants, and in particular, for $N > 1$, the relations, 

\begin{equation}
A_i \ll  \eta \rr = C_i \ll  \eta \rr = I, \quad 
C_i \ll -\eta \rr = D_i \ll -\eta \rr = R_a, 
\end{equation}

\noindent still apply, where $\eta$ is the crossing parameter, and $R_a$ is the diagonal matrix
with entries,  

\begin{equation}
\ll R_a \rr_{h, \, h^{\, \prime} } = [a]^{1/2} \, \delta \ll \h, \, \h^{\, \prime} \rr, 
\end{equation}

\noindent and $a$ is the fixed height at the site at the centre of the lattice. Following 
section \textbf{\ref{evaluating.the.local.height.probabilities}}, we write, 

\begin{equation}
\prod_i A_i \ll \eta \rr  \, B_i \ll - \eta \rr  \, C_i \ll \eta \rr  \, D_i \ll - \eta \rr  = R_a^{\, 2N}, 
\quad 
M_{1, 1} \, M_{2, 1} \cdots M_{3, N} \, M_{4, N} = R_a^{\, 2N} \, e^{\, - 4 \, N \, \eta \, \H}, 
\end{equation}

\noindent and we obtain, 

\begin{equation}
P_N \ll a, b, b \pm 1 \, \vert \, p_N \rr =
\frac{
\textit{Trace} 
\ll S_a \, R_a^{\, 2N} \, e^{\, -4 \, N \, \eta \, \H} \rr}{     
\textit{Trace} 
\ll        R_a^{\, 2N} \, e^{\, -4 \, N \, \eta \, \H} \rr
}
\end{equation}

\subsubsection{Step 3 and Step 4 on a spiral}
These two steps carry through without modification from a plane to a deep spiral.

\subsubsection{Step 5 on a spiral}
Working in terms of elliptic theta functions with a conjugate nome $q_{\, N}$, that depends
on the winding number $N$, the arguments 
of Andrews \textit{et al.} carry through with minimal modification and we obtain, 

\begin{equation}
\ll e^{\, - \, 4 \, N \, \eta \, \H} \, \rr_{h, \, h^{\, \prime}} = 
q_{\, N}^{\, N \ll \Phi \ll h \rr - \ll 2 a - m - 1 \rr^2/16 \ll m + 1 \rr \rr} \, 
\delta \ll \h, \, \h^{\, \prime} \rr,
\label{another.exponential}
\end{equation}

\noindent where $\Phi$ is the same $L \rightarrow \infty$ limit of the corner transfer matrix 
energy function $\Phi_{\, n}$ in (\ref{corner.transfer.matrix.energy.function}), and we also 
obtain, 

\begin{equation}
\ll R_a                          \rr_{h, \, h^{\, \prime}} = 
\tau^{\, N} \, q_{\, N}^{\, N \ll a  - m-1 \rr^2 / 8 \ll m+1 \rr} E^{\, N} \ll q_{\, N}^{\, a / 2}, y \rr \, 
\delta \ll \h, \, \h^{\, \prime}  \rr
\end{equation}

\noindent Using these results,

\begin{equation}
P_N \ll a, b, b \pm 1 \, | \, q_{\, N} \rr =
\frac{
E^{\, N} \ll q_{\, N}^{\, a/2}, q_{\, N}^{\, \ll m + 1 \rr/2} \rr \, 
X        \ll a, b, b \pm 1 \, \vert \, q_{\, N}^{\, N} \rr
}{
\sum_{1 \leq \, a^{\, \prime} \, < \, r}
E^{\, N} \ll q_{\, N}^{\, a^{\, \prime} / 2}, q_{\, N}^{\, \ll m + 1 \rr/2} \rr \, 
X        \ll       a^{\, \prime}, \, b, \, b \pm 1 \, \vert \, q_{\, N}^{\, N} \rr
}, 
\label{P.N}
\end{equation}

\noindent where the one-dimensional configuration sum 
$X \ll a, b, b \pm 1 \, \vert \, q_{\, N}^{\, N} \rr$ is a $q_{\, N}^{\, N}$-series. 

\subsubsection{Remark} It is crucial to note that the one and only effect of the winding 
number $N$ on the essential factor in the corner transfer matrix computation of the local 
height probability is the factor $N$ in the exponential in (\ref{another.exponential}), and 
that this factor simply modifies the modular parameter 
as $\tau_{\, N} \rightarrow N \, \tau_{\, N}$. This is the essential observation in this work.

\subsubsection{Step 6 on a spiral} The one-dimensional configuration sum 
$X \ll a, b, b \pm 1 \, \vert \, q_{\, N}^{\, N} \rr$ satisfies the same difference equations, 
and the same initial conditions as 
$X \ll a, b, b \pm 1 \, \vert \, q        \rr$, and can be evaluted similarly. 
The result is proportional to the same Virasoro character as before, but with
$q \rightarrow q_{\, N}^{\, N}$.

\subsection{Scaling $L$ and $\tau$ with $N$}
In the sequel, as we vary the number of quadrants $4N$, we need to scale $L$, the number 
of sites in a one-dimensional configuration, which is basically a measure of the radial 
extension of the lattice, and $\tau$, which is a measure of departure from criticality, 
as functions of $N$. We call the scaled parameters $L_{\, N}$ and $\tau_{\, N}$, and 
choose them as,

\begin{equation}
L_N = \ll N L \rr^{\, N}, \quad
\tau_{\, N} = \tau \, / \, N, 
\end{equation}

\noindent where $L = L_{1}$ and $\tau = \tau_{1}$, the parameters that we start from on 
a plane. 

\subsubsection{Remark} We interpret the multiplicative factor $N$ in $\ll N L \rr^{\, N}$ 
as a filling in of the lattice with more vertices, so that the limit $N \rightarrow \infinity$, 
can be interpreted as a thermodynamic limit, and at criticality, we end up with a system of finite 
size $\ll L^{\, \prime} \rr^{\, N}$, where $L^{\, \prime}$ is a finite length of a system with 
continuous degrees of freedom. The power $N$ further increases the size of the lattice, but will 
be undone as we conformally transform the spiral geometry to an annular geometry.

\subsection{A change of normalization}
Consider (\ref{P.N}). Since we wish to compute critical partition functions, 
rather than local height probabilities, we are allowed to choose convenient normalization 
such that the quantities that we are computing do not necessary add up to 1, as probabilities 
do. We need to choose a normalization that agrees with (\ref{P}) at $N=1$, but 
is allowed to be different for $N\neq 1$. We consider the quantity,

\begin{equation}
P_N^{\, \prime} \ll a, b, b \pm 1 \, | \, q_{\, N} \rr =
\frac{
E^{\, N} \ll q_{\, N}^{\, a / 2}, \, q_{\, N}^{\, \ll m + 1 \rr / 2} \rr
X        \ll a, \, b, \, b \pm 1 \, \vert \, q_{\, N}^{\, N} \rr}{
\sum_{1 \, \leq \, a^{\, \prime} \, < \, r} 
\ll 
E   \ll q_{\, N}^{\, a^{\, \prime} / 2}, \, q_{\, N}^{\, \ll m + 1 \rr / 2} \rr 
X   \ll       a^{\, \prime},     \, b, \, b \pm 1 \, \vert \, q_{\, N}^{\, N} \rr 
\rr^{\, N}
}
\label{P.N.normalized.differently}
\end{equation}

\noindent We need this modified normalization to make contact with the results of 
\cite{saleur.bauer}. Notice that the new normalization differs from the original 
by an overall factor that does not change the relative weights of the configurations.

\subsection{Writing the local height probabilities on a spiral in terms of affine 
and Virasoro characters}

On a spiral, (\ref{rewriting}) becomes,  

\begin{multline}
P_N \ll a, b, b \pm 1 \, | \, q_{\, N} \rr = 
\ll
\frac{
\Theta_{\, j_{\, 3}, \, m-1} \ll q_{\, N}^{\, 1/2}, q_{\, N} \rr 
\Theta_{\,        1, \,   2} \ll q_{\, N}^{\, 1/2}, q_{\, N} \rr }{
\Theta_{\, j_{\, 1}, \,   m} \ll q_{\, N}^{\, 1/2}, q_{\, N} \rr 
\Theta_{\, j_{\, 3}, \,   3} \ll q_{\, N}^{\, 1/2}, q_{\, N} \rr 
} 
q_{\, N}^{\, -\nu}
\rr^{\, N} 
\ll q_{\, N}^{\, N \, \nu} c_{\, j_{\, 1}\, j_{\, 2}\, j_{\, 3}} \ll q_{\, N}^{\, N} \rr \rr \\
=
\ll
\frac{
\Theta_{\, j_{\, 1},     m-1} \ll q_{\, N}^{\, 1/2}, q_{\, N} \rr 
\Theta_{\,        1,   \,  2} \ll q_{\, N}^{\, 1/2}, q_{\, N} \rr
}{
\Theta_{\, j_{\, 2},       m} \ll q_{\, N}^{\, 1/2}, q_{\, N} \rr 
\Theta_{\, j_{\, 2},   \,  3} \ll q_{\, N}^{\, 1/2}, q_{\, N} \rr
} 
\rr^{\, N} 
c_{\, j_{\, 1}\, j_{\, 2}\, j_{\, 3}} \ll q_{\, N}^{\, N} \rr,  
\label{P.N.Virasoro}
\end{multline}

\noindent where the parameters $a, b,$ and $b \pm 1$ on the left are related to the parameters 
$j_{\, 1}, j_{\, 2},$ and $j_{\, 3}$ on the right as in (\ref{notation.01}--\ref{notation.02}). 
Finally, we formally take $N \in \RR$. Equation (\ref{P.N.Virasoro}), with $N \in \RR$, 
is the result that we need.

\section{Critical partition functions on a cylindrical geometry revisited}
\label{section.cylinder.geometry}

\subsection{Two regularizations before taking the infinite-spiral limit}
\label{step.02}
To take the $N \rightarrow \infty$ of a critical theory on an infinite-spiral, and more importantly, 
to conformally transform the spiral to an annulus with a finite aspect-ratio that can be conformally 
transformed to a cylinder, so that we end up with a critical model on a cylinder, we need two 
regularizations.

\subsubsection{Regularization 1. Infrared} 
We take the radial extension of the spiral to be a large radius $r_{outer}$, rather 
than $\infty$, which is an \lq infrared-type\rq\ regularization that allows us to 
compute on a finite lattice, then take the thermodynamic limit. This is how the local 
height probabilities are evaluated, following \cite{andrews.baxter.forrester}.

\subsubsection{Regularization 2. Ultraviolet} 
We remove an inner spiral of small radius $r_{inner}$, which is an \lq ultraviolet-type\rq\ 
regularization that allows us to avoid a potential singularity near the origin in subsequent 
computations. One can motivate this regularization as follows. 
Think of the vertices on which the height variables are located as infinitesimal punctures.
The height variable assigned to each vertex is distributed on the circumference. 
We can actually take the radii of these punctures to be finite, rather than infinitesimal. 
In fact, we can take the radii of different punctures to be all different, and show that 
the Yang-Baxter equations remain unaffected. What we wish to do here is to keep all radii 
infinitesimal, but take that of the vertex at the origin to be finite and small.
We end up with a lattice that has \textit{\lq a hole in the middle\rq}, as in 
Figure \textbf{\ref{figure.puncture}}. 
The height variable at the center of the lattice is now distributed on the finite 
circumference.
In a $4N$-quadrant geometry, the circumference of the puncture at the origin becomes 
a finite-radius spiral, with the top and the bottom end-points identified. The small 
radius $r_{inner}$ that we are interested in is the radius of this inner spiral. One can 
now think of taking the continuum limit as increasing the number of vertices in the lattice, 
while keeping $r_{inner}$ and $r_{outer}$ fixed. 

\begin{figure}
\begin{tikzpicture}[scale=.6]


\draw [thick] (0, 5) rectangle (1,6);

\draw [thick] (0, 4) rectangle (1,5);
\draw [thick] (1, 4) rectangle (2,5);

\draw [thick] (0, 3) rectangle (1,4);
\draw [thick] (1, 3) rectangle (2,4);
\draw [thick] (2, 3) rectangle (3,4);

\draw [thick] (0, 2) rectangle (1,3);
\draw [thick] (1, 2) rectangle (2,3);
\draw [thick] (2, 2) rectangle (3,3);
\draw [thick] (3, 2) rectangle (4,3);

\draw [thick] (0, 1) rectangle (1,2);
\draw [thick] (1, 1) rectangle (2,2);
\draw [thick] (2, 1) rectangle (3,2);
\draw [thick] (3, 1) rectangle (4,2);
\draw [thick] (4, 1) rectangle (5,2);

\draw [very thick] (-0.025,-7.0)--(-0.025,7.0);
\draw [very thick] ( 0.025,-7.0)--( 0.025,7.0);

\draw [very thick] (-7.0, -0.025)--(0.0, -0.025);
\draw [very thick] (-7.0,  0.025)--(0.0,  0.025);

\draw  [very thick] (0, 0)--(1,  0.25); 
\draw  [very thick] (0, 0)--(1, -0.25);

\draw  [very thick] (1,  0.25)--(7,  0.25);
\draw  [very thick] (1, -0.25)--(7, -0.25);

\draw  [thick] (5,  1)--(6,  1);
\draw  [thick] (5, -1)--(6, -1);

\draw  [thick] (1,  1)--(1, 0.25); 
\draw  [thick] (2,  1)--(2, 0.25); 
\draw  [thick] (3,  1)--(3, 0.25); 
\draw  [thick] (4,  1)--(4, 0.25); 
\draw  [thick] (5,  1)--(5, 0.25); 
\draw  [thick] (6,  1)--(6, 0.25); 

\draw  [thick] (1,  -1)--(1, -0.25); 
\draw  [thick] (2,  -1)--(2, -0.25); 
\draw  [thick] (3,  -1)--(3, -0.25); 
\draw  [thick] (4,  -1)--(4, -0.25); 
\draw  [thick] (5,  -1)--(5, -0.25); 
\draw  [thick] (6,  -1)--(6, -0.25);

\draw [thick] (0, -2) rectangle (1,-1);
\draw [thick] (1, -2) rectangle (2,-1);
\draw [thick] (2, -2) rectangle (3,-1);
\draw [thick] (3, -2) rectangle (4,-1);
\draw [thick] (4, -2) rectangle (5,-1);

\draw [thick] (0, -3) rectangle (1,-2);
\draw [thick] (1, -3) rectangle (2,-2);
\draw [thick] (2, -3) rectangle (3,-2);
\draw [thick] (3, -3) rectangle (4,-2);

\draw [thick] (0, -4) rectangle (1,-3);
\draw [thick] (1, -4) rectangle (2,-3);
\draw [thick] (2, -4) rectangle (3,-3);

\draw [thick] (0, -5) rectangle (1,-4);
\draw [thick] (1, -5) rectangle (2,-4);

\draw [thick] (0, -6) rectangle (1,-5);


\draw [thick] ( 0, 5) rectangle (-1,6);

\draw [thick] ( 0, 4) rectangle (-1,5);
\draw [thick] (-1, 4) rectangle (-2,5);

\draw [thick] ( 0, 3) rectangle (-1,4);
\draw [thick] (-1, 3) rectangle (-2,4);
\draw [thick] (-2, 3) rectangle (-3,4);

\draw [thick] ( 0, 2) rectangle (-1,3);
\draw [thick] (-1, 2) rectangle (-2,3);
\draw [thick] (-2, 2) rectangle (-3,3);
\draw [thick] (-3, 2) rectangle (-4,3);

\draw [thick] ( 0, 1) rectangle (-1,2);
\draw [thick] (-1, 1) rectangle (-2,2);
\draw [thick] (-2, 1) rectangle (-3,2);
\draw [thick] (-3, 1) rectangle (-4,2);
\draw [thick] (-4, 1) rectangle (-5,2);

\draw [thick] ( 0, 0) rectangle (-1,1);
\draw [thick] (-1, 0) rectangle (-2,1);
\draw [thick] (-2, 0) rectangle (-3,1);
\draw [thick] (-3, 0) rectangle (-4,1);
\draw [thick] (-4, 0) rectangle (-5,1);
\draw [thick] (-5, 0) rectangle (-6,1);

\draw [thick] ( 0, -1) rectangle (-1,0);
\draw [thick] (-1, -1) rectangle (-2,0);
\draw [thick] (-2, -1) rectangle (-3,0);
\draw [thick] (-3, -1) rectangle (-4,0);
\draw [thick] (-4, -1) rectangle (-5,0);
\draw [thick] (-5, -1) rectangle (-6,0);

\draw [thick] ( 0, -2) rectangle (-1,-1);
\draw [thick] (-1, -2) rectangle (-2,-1);
\draw [thick] (-2, -2) rectangle (-3,-1);
\draw [thick] (-3, -2) rectangle (-4,-1);
\draw [thick] (-4, -2) rectangle (-5,-1);

\draw [thick] ( 0, -3) rectangle (-1,-2);
\draw [thick] (-1, -3) rectangle (-2,-2);
\draw [thick] (-2, -3) rectangle (-3,-2);
\draw [thick] (-3, -3) rectangle (-4,-2);

\draw [thick] ( 0, -4) rectangle (-1,-3);
\draw [thick] (-1, -4) rectangle (-2,-3);
\draw [thick] (-2, -4) rectangle (-3,-3);

\draw [thick] ( 0, -5) rectangle (-1,-4);
\draw [thick] (-1, -5) rectangle (-2,-4);

\draw [thick] ( 0, -6) rectangle (-1,-5);

\foreach \x in {0,...,12}
{
\draw [fill=black!50] (\x-6, 1) circle (0.12);
\draw [fill=black!50] (\x-6,-1) circle (0.12);
}

\foreach \x in {0,...,7}
{
\draw [fill=black!50] (\x-7, 0) circle (0.12);
}

\foreach \x in {7,...,13}
{
\draw [fill=black!50] (\x-6, 0.24) circle (0.12);
\draw [fill=black!50] (\x-6,-0.24) circle (0.12);
}

\foreach \x in {0,...,10}
{
\draw [fill=black!50] (\x-5, 2) circle (0.12);
\draw [fill=black!50] (\x-5,-2) circle (0.12);
}

\foreach \x in {0,...,8}
{
\draw [fill=black!50] (\x-4, 3) circle (0.12);
\draw [fill=black!50] (\x-4,-3) circle (0.12);
}

\foreach \x in {0,...,6}
{
\draw [fill=black!50] (\x-3, 4) circle (0.12);
\draw [fill=black!50] (\x-3,-4) circle (0.12);
}

\foreach \x in {0,...,4}
{
\draw [fill=black!50] (\x-2, 5) circle (0.12);
\draw [fill=black!50] (\x-2,-5) circle (0.12);
}

\foreach \x in {0,...,2}
{
\draw [fill=black!50] (\x-1, 6) circle (0.12);
\draw [fill=black!50] (\x-1,-6) circle (0.12);
}

\draw [fill=black!50] (0, 7) circle (0.12);
\draw [fill=black!50] (0,-7) circle (0.12);

\node at ( 4.5, 4.5) {$\cA_{\, i}$};
\node at (-4.5, 4.5) {$\cB_{\, i}$};
\node at (-4.5,-4.5) {$\cC_{\, i}$};
\node at ( 4.5,-4.5) {$\cD_{\, i}$};

\draw [fill=black!50] (0.00, 0.00) circle (0.50);

\end{tikzpicture}
\caption{
\textit{
}
}
\label{figure.puncture}
\end{figure}

\subsection{From a $4N$-quadrant to an infinite spiral}

From (\ref{P.N.normalized.differently}), the one-dimensional sum in 
the numerator in the local height probability depends on $q_N$ and $N$ only through 
the combination $q_{\, N}^{\, N} = e^{\, -2 \, \pi \, N \, \tau_N}$.  We require 
that, 

\begin{equation}
N           \rightarrow \infty, 
\quad
\tau_{\, N} \rightarrow 0, 
\quad
N \, \tau_N = \tau_{\, 0} > 0, 
\quad
L_{\, N} \rightarrow \ll N \, L \rr^{\, N} = \ll L^{\, \prime} \rr^{\, N},
\label{large.N.limit.02}
\end{equation}

\noindent where $\tau_{\, 0}$ is a finite real number, and $L^{\, \prime}$ is a finite 
length in a theory with continuum degrees of freedom
\footnote{\,
We understand the limit 
$N           \rightarrow \infty$ as $N           \rightarrow    N_{\textit{very large}}$, 
$\tau_{\, N} \rightarrow      0$ as $\tau_{\, N} \rightarrow \tau_{\textit{very small}}$,
where $N_{\textit{very large}}$ and $\tau_{\textit{very small}}$ are very large and very 
small numbers, respectively, compute, then formally take the limits 
$N_{\textit{very large}}    \rightarrow \infty$, and 
$\tau_{\textit{very small}} \rightarrow 0$
}.
In this limit, the system is asymptotically close to criticality, on a spiral with 
$4N$-quadrants, an inner radius $r_{inner}$, and an outer radius $r_{outer}$. In the sequel, 
we choose to measure distances in the theory with continuum degrees of freedom such that,

\begin{equation}
r_{inner} = 1, 
\quad
r_{outer} = \ll L^{\, \prime} \rr^{\, N}
\end{equation}

\noindent Choosing $\tau_{\, N} = \tau / N$, the one-dimensional configuration sum (the second 
factor) in the numerator of (\ref{P.N.normalized.differently}) is independent of $N$. 
In the limit (\ref{large.N.limit.02}), this factor remains invariant, but the first factor, as 
well as the denominator diverge because working in terms of the nome $q_{\, N}$ is inadequate 
in the critical limit $q_{\, N} \rightarrow 1$. 
To deal with that, we need to perform a conjugate modulus transformation and work in terms of 
$p_{\, N} = e^{\, - 2 \, \pi / \tau_{\, N}}$. Using the identity
\footnote{\,
Equation (4.4 c) in \cite{date.jimbo.kuniba.miwa.okado.03}, with appropriate changes
in notation. 
}, 

\begin{equation}
\Theta_{\, k, \, l} \ll q^{\, \frac12}, \, q \rr = 
q^{\, - \, l \, / \, 4} \, 
\ll \frac{\ll m + 1 \rr}{  l \, \tau } \rr^{\, \frac12} \,  
\theta_{\, 1}      \, \ll \pi \, k / l, \, p^{\, \ll m + 1 \rr / \, l} \rr, 
\end{equation}

\noindent where the nome $p$, and the conjugate nome $q$ are defined in (\ref{modular.parameter.conjugation}), 
(\ref{P.N.Virasoro}) can be re-written as, 

\begin{equation}
\ll
\ll \frac{3 m}{2 \ll m + 1 \rr} \rr^{\, 1/2}
\frac{
\theta_{\, 1} \ll \pi \, j_{\, 3} / \ll m + 1 \rr, p_{\, N}^{\, 2 / \ll m + 1 \rr} \rr 
\theta_{\, 1} \ll \pi        /     2,         p_{\, N}                        \rr
}
{
\theta_{\, 1} \ll \pi \, j_{\, 1}  / m, p_{\, N}^{\, 2/m} \rr
\theta_{\, 1} \ll \pi \, j_{\, 2}  / 3, p_{\, N}^{\, 2/3} \rr
}
\rr^{\, N} 
c_{\, j_{\, 1} \, j_{\, 2} \, j_{\, 3}} \ll q_N^{\, N} \rr 
\end{equation}

\noindent In the critical limit, this reduces to,

\begin{equation}
\ll
\ll \frac{3m}{2 \ll m+1 \rr} \rr^{\, 1/2}
\frac{
\sin \ll \pi j_{\, 3} / \ll m+1 \rr \rr}{
\sin \ll \pi j_{\, 1} /     m   \rr \sin \ll \pi j_{\, 2} /3\rr
}
\rr^{\, N} 
c_{j_{\, 1} j_{\, 2} j_{\, 3}} \ll q_{\, N}^{\, N} \rr 
\sim
c_{j_{\, 1} j_{\, 2} j_{\, 3}} \ll q \rr 
= 
\chi^{\, Vir}_{\, r, s} \ll q \rr,  
\label{P.criticality}
\end{equation}

\noindent 
In the first proportionality, we have dropped a constant multiplicative factor, 
and used $q_N = q^{\, 1 / N}$, and 
in the second equality,       we have used (\ref{notation.01}--\ref{notation.02}). 

\subsection{Re-interpretation of $\tau$}

Equation (\ref{P.criticality}) describes a partition function in a critical model in 
a specific geometry, with specific
boundary conditions. At criticality, the effective nome $q_{\, N}^{\, N} = q$ that appears 
in (\ref{P.criticality}) must be re-interpreted. Its logarithm can no longer be considered as 
measuring departure from criticality. From experience with calculations in finite geometries
\cite{saleur.bauer}, we expect it to be related to ratios of the dimensions of the system. 
In that case, it should remain invariant under conformal transformations.

\subsection{Conformal transformation 1. From an infinite spiral to an annulus}

Now that we are (asymptotically) close to criticality, the system is conformally invariant, 
and we can use conformal transformations to bring it back to a familiar geometry. Using the 
conformal transformation

\begin{equation}
z \rightarrow z^{\, \frac{1}{N}} = \ll r e^{\, i \theta} \rr^{\, \frac{1}{N}}, 
\end{equation}

\noindent takes us from the spiral geometry to an annulus. The ratio 
$\frac{r_{outer}}{r_{inner}} = \ll L^{\, \prime} \rr^{\, N}$ which determines the radial 
extension of the deep spiral transforms as, 

\begin{equation}
\ll L^{\, \prime} \rr^{\, N} \rightarrow L^{\, \prime}, 
\end{equation}

\noindent and we end up on an annulus that has an inner radius $r_{\, \textit{inner}} = 1$, 
and an outer radius $r_{\, \textit{outer}} = L^{\, \prime}$. 

\subsection{Conformal transformation 2. From an annulus to a cylinder}

We map the annulus to a cylinder of width $T$ in the periodic direction, and height 
$L^{\, \prime}$ in the non-periodic direction, using the conformal transformation, 

\begin{equation}
z \rightarrow w = \textit{log \,} z, 
\end{equation}

\noindent which puts us on a strip with,  

\begin{equation}
2 \pi H = \textit{log \,} L^{\, \prime}, \quad T = 2 \pi, 
\end{equation}

\noindent where the factor $2 \pi$ in $2 \pi H$ is chosen to simplify the notation in the sequel.
Since we have maintained periodic boundary conditions in the angular direction, this strip is 
equivalent to a cylinder of aspect ratio $H$. The finite size that appears in the computations 
of the characters, as given in \cite{saleur.bauer} is parameterized by the nome 
$q_N^{\, N} = q = e^{\, - \, 4 \, \pi \, \tau / \ll m+1 \rr}$, where 
$\tau = T / \ll 2 \pi H \rr$. In our case, 

\begin{equation}
\tau = \frac{2 \pi}{ 2 \pi H} = \frac{1}{H}, 
\end{equation}

\noindent in complete agreement with the result of Saleur and Bauer \cite{saleur.bauer}. 
The factor 
raised to a power $N$ in (\ref{}) is such that, under the first conformal transformation, 
this power is canceled, and we end up with a finite numerical factor that, for the purposes of this work, 
can be ignored.

\section{Postscript}
\label{section.postscript.comments}

\subsection{Surface critical exponents in cyclic solid-on-solid models}
\label{work.seaton.nienhuis}
In \cite{seaton.nienhuis}, Seaton and Nienhuis used the spiral geometry
that interpolates off-critical models on a plane and critical models on a cylinder,
as proposed in \cite{foda} and reproduced in the present work, to compute surface critical 
exponents in cyclic solid-on-solid models.

\subsection{$q$-Virasoro algebra} 
\label{work.shiraishi.kubo.awata.odake}
In \cite{shiraishi.kubo.awata.odake}, Shiraishi, Kubo, Awata and Odaka obtained a $q$-deformed 
version of the Virasoro algebra that extends the known Virasoro algebra off-criticality. 
The identification in \cite{date.jimbo.miwa.okado.01} of Virasoro characters as the essential 
factors in Regime-III local height probabilities was surprising when first discussed in the 
1980's because, at the time, such a deformation was not known. Now that we know that this 
deformation exists, the appearance of Virasoro characters off-criticality, should be less 
of a surprise. The interpolation obtained in this paper is a concrete way 
to see how the same object, $ X\ll a \, | \,  b, \, b \pm 1 \, \vert \, q \rr$ appears 
at criticality as a partition function on a cylinder, and also off-criticality as the essential 
factor in a local height probability.

\subsection{Two ways to introduce a scale lead to identical results}
A critical system is characterized by the absence of a characteristic mass 
or length scale. A characteristic scale can be introduced either 
\textbf{ 1.} By varying a physical parameter such as the temperature away 
from its critical value, or 
\textbf{ 2.} By introducing finite size effects. The effects of \textbf{1} and \textbf{2} 
on the same initial critical system are physically distinct. However, we know 
from \cite{date.jimbo.miwa.okado.01, date.jimbo.miwa.okado.02, 
saleur.bauer} that the off-critical one-point functions in regime-III restricted 
solid-on-solid models \cite{andrews.baxter.forrester}, and 
the partitions functions of the corresponding minimal conformal field theories 
on a cylinder, can be expressed (up to an overall normalization) in terms of the 
same characters of Virasoro degenerate irreducible highest-weight modules, which 
are $q$-series in an indeterminate $q = e^{\, - \, 4 \, \pi \, \tau \, / \, \ll m+1 \rr}$. 
Off criticality, on a plane,  $\tau$ is a parameter that measures departure from criticality. 
At  criticality, on a cylinder, $\tau$ is the aspect-ratio (circumference to width) of the cylinder
\footnote{\, 
As in \cite{date.jimbo.miwa.okado.01, date.jimbo.miwa.okado.02, saleur.bauer}, 
we restrict our attention in the present work to regime-III restricted solid-on-solid models and the
corresponding minimal conformal field theories. Our result is expected to extend to models other
solid-on-solid models. Following the appearance of the first version of this work \cite{foda}, it was 
extended in \cite{seaton.nienhuis} to cyclic solid-on-solid models.
}. 

\subsection{Interpretation. An interplay of geometry and physics}
One way to understand the above result is to say that, if we ignore normalization
factors that do not carry information about the statistical mechanics, 
the essential part of the the local height probability 
$P \ll a \, | \,  b, \, b \pm 1 \, \vert \, q \rr$, namely the one-dimensional sum 
$X \ll a \, | \,  b, \, b \pm 1 \, \vert \, q \rr$ is a mathematical object that 
can be interpreted in two different ways, 
\1 as an off-critical one-point function on a plane, $\tau$-away from criticality, and 
\2 as a critical partition function on a cylinder of aspect-ratio $\tau$. 
The fact that the same $q$-series appears in these two physically-distinct settings 
points to an interplay between the geometry of the surface that a model lives on, and 
the physics of the model, in the sense of how far it is from criticality. To understand 
this interplay better, we extend the definition of the off-critical local height probabilities 
from the 4-quadrant plane geometry to a $4N$-quadrant spiral geometry, then formally take 
$N \in \RR$, to obtain an object that interpolates the off-critical local height probabilities 
on a plane and the partition functions on a finite cylinder which are Virasoro characters.

\subsubsection{Constant-$\ll N\, \tau \rr$ curves}
Consider the parameter space with coordinates $\ll N, \tau \rr$, where $1 \leq \,  N < \, \infty$ 
parameterizes the geometry of the configuration that the model lives on, and $0 < \, \tau < \, \infty$ 
parameterizes its departure from criticality. 
Our result is that this parameter spaces is foliated by constant-$\ll N\, \tau \rr$ curves, 
$N \, \tau = \tau_{\, 0} \in \RR$, that we call $\tau_{\, 0}$-curves, where $0 < \tau_{\, 0} < \infty$ 
is constant, and the local height probability is invariant, as a $q$-series, on a $\tau_{\, 0}$-curve. 
At the $\ll 1, \tau \rr$-end of a $\tau_{\, 0}$-curve, the $q$-series has the interpretation of 
a local height probability. 
At the $\ll N \rightarrow \infty, \tau \rightarrow 0 \rr$-end, it has the interpretation of 
critical partition function on a cylinder, which following Saleur and Bauer is a Virasoro 
character \cite{saleur.bauer}. 
As $q$-series, for $\tau = - \, \textit{log} \, q = \tau_{\, 0}$, these two limits are equal. 

\subsection{Epilogue} 
The starting point of this work is an observation by Date, Jimbo, Miwa and Okada
\cite{date.jimbo.miwa.okado.01}, followed by another observation by Saleur and Bauer 
\cite{saleur.bauer}. 
In \cite{date.jimbo.miwa.okado.02}, Date \textit{et al.} conclude a review of their work 
on the subject with 
\textit{\lq 
Modular invariance plays two different roles in the 2D lattice systems. Are they totally 
unrelated to each other? It is unbelievable. Then what is the structure governing both? 
We stop the lecture with this question.\rq} 
In this work, we have argued that local height probabilities on a spiral geometry are interpolating 
objects, with well-defined modular properties, that reduce to each of the objects referred to by Date
\textit{et al.} in appropriate limits. We wish to argue that the existence of these interpolating 
objects, and their functional dependence on the winding number $N$ and departure from criticality 
$\tau$, realizes the relation forecast by Date \textit{et al.} 

\section*{Acknowledgements}
I would like to thank 
H Saleur  for a discussion on \cite{saleur.bauer} in 1988, and 
J B Zuber for a discussion on \cite{cardy.02} in 2017. This work was supported by the Australian 
Research Council.


\begin{thebibliography}{99}

\bibitem{baxter.1980}
R J Baxter,
\textit{Hard hexagons: exact solution}, 
Journal of Physics \textbf{A}: Mathematical and General \textbf{13.3} (1980) L61--L70

\bibitem{baxter.1981}
R J Baxter,
\textit{Roger-Ramanujan Identities in the hard Hexagon Model}, 
Journal of Statistical Physics \textbf{26.3} (1981) 427--452

\bibitem{andrews.baxter.forrester}
G E Andrews, R J Baxter, and P J Forrester,
\textit{Eight-vertex SOS model and generalized Rogers-Ramanujan-type identities}, 
Journal of Statistical Physics \textbf{35.3} (1984) 193--266

\bibitem{gradshteyn.ryzhik}
I S Gradshteyn and I M Ryzhik,
\textit{Table of integrals, series, and products},
Academic Press (1994),
\texttt{ISBN-10: 012294755X, ISBN-13: 978-0122947551}

\bibitem{friedan.qiu.shenker.01}
D Friedan, Z Qiu, and S Shenker,
\textit{Conformal invariance, unitarity, and critical exponents in two dimensions},
Physical Review Letters \textbf{52.18} (1984) 1575--1578

\bibitem{friedan.qiu.shenker.02}
D Friedan, Z Qiu, and S Shenker,
\textit{Conformal invariance, unitarity and two dimensional critical exponents},
419--449, 
in \textit{Vertex operators in mathematics and physics}, 
Mathematical Sciences Research Institute Publications (Book 3),
J Lepowsky (Editor), S Mandelstam (Editor), and I Singer (Editor),
Springer, New York, NY (1985) 
\texttt{ISBN 0387961216}

\bibitem{belavin.polyakov.zamolodchikov.01}
A A Belavin, A M Polyakov, and A B Zamolodchikov,
\textit{Infinite conformal symmetry of critical fluctuations in two dimensions},
Journal of Statistical Physics \textbf{34.5} (1984) 763--774

\bibitem{belavin.polyakov.zamolodchikov.02}
A A Belavin, A M Polyakov, and A B Zamolodchikov,
\textit{Infinite conformal symmetry in two-dimensional quantum field theory},
Nuclear Physics \textbf{B 241} (1984) 333--380

\bibitem{forrester.baxter}
P J Forrester and R J Baxter, 
\textit{Further exact solutions of the eight-vertex SOS model and generalizations 
of the Rogers-Ramanujan identities}, 
Journal of Statistical Physics \textbf{38.3} (1985) 435--472

\bibitem{pasquier.01}
V Pasquier, 
\textit{Two-dimensional critical systems labelled by Dynkin diagrams},
Nuclear Physics \textbf{B 285} (1987) 162--172

\bibitem{pasquier.02}
V Pasquier, 
\textit{Exact solubility of the $D_{\, n}$ series},
Journal of Physics\textbf{A}: Mathematical and General \textbf{20.4} (1987) L217--L220 

\bibitem{pasquier.03}
V Pasquier, 
\textit{$D_{\, n}$ models: local densities},
Journal of Physics \textbf{A}: Mathematical and General \textbf{20.4} (1987) L221--L226 

\bibitem{itzykson.saleur.zuber.book}
C Itzykson, H Saleur, and J- Zuber, Editors,
\textit{Conformal invariance and applications to statistical mechanics},
World Scientific Publishing Company (1998) Singapore, 
\texttt{ISBN-10: 9971506068, ISBN-13: 978-9971506063}

\bibitem{pasquier.04}
V Pasquier, 
\textit{Lattice derivation of modular invariant partition functions on the torus},
Journal of Physics \textbf{A}: Mathematical and General \textbf{20.18} (1987) L1229--L1237 

\bibitem{nienhuis.01}
B Nienhuis, 
\textit{Critical behavior of two-dimensional spin models and charge asymmetry in 
the Coulomb gas},
Journal of Statistical Physics \textbf{34.5-6} (1984) 731--761

\bibitem{nienhuis.02}
B Nienhuis, 
\textit{Two-dimensional critical phenomena and the Coulomb Gas}, 
in 
\textit{Phase Transitions and Critical Phenomena}, 
C Domb, M Green, and J L Lebowitz, Editors, volume \textbf{11} (1987) Academic Press,
\texttt{ISBN-13: 978-0122203114, ISBN-10: 0122203119}

\bibitem{kuniba.yajima.01}
A Kuniba and T Yajima,
\textit{Local state probabilities for an infinite sequence of solvable lattice models},
Journal of Physics \textbf{A}: Mathematical and General \textbf{21.2} (1988) 519--527

\bibitem{kuniba.yajima.02}
A Kuniba and T Yajima,
\textit{Local state probabilities for solvable restricted solid-on-solid models: 
$A_n$, $D_n$, $D_n^{(1)}$, and $A_n^{(1)}$},
Journal of statistical physics \textbf{52.3--4} (1988) 829--883

\bibitem{pearce.seaton.01}
P A Pearce and K A Seaton,
\textit{Solvable hierarchy of cyclic solid-on-solid lattice models},
Physical Review Letters \textbf{60.14} (1988) 1347--1350

\bibitem{pearce.seaton.02}
P A Pearce K A Seaton,
\textit{Exact solution of cyclic solid-on-solid lattice models},
Annals of Physics \textbf{193.2} (1989) 326--366

\bibitem{baxter.1968}
R J Baxter,
\textit{Dimers on a rectangular lattice},
Journal of Mathematical Physics \textbf{9.4} (1968) 650--654

\bibitem{baxter.1978}
R J Baxter,
\textit{Variational approximations for square lattice models in statistical mechanics},
Journal of Statistical Physics \textbf{19.5} (1978) 461--478

\bibitem{baxter.review}
R J Baxter,
\textit{Corner transfer matrices},
Physica \textbf{106A} (1981) 18--27

\bibitem{baxter.book}
R J Baxter,
\textit{Exactly Solved Models in Statistical Mechanics},
Academic Press, 1982,
\texttt{ISBN 0120831805}

\bibitem{andrews.1981}
G E Andrews,
\textit{The hard-hexagon model and Rogers-Ramanujan type identities}, 
Proceedings of the National Academy of Sciences USA \textbf{78.9} (1981) 5290--5292

\bibitem{rocha.caridi}
A Rocha-Caridi,
\textit{Vacuum vector representations of the Virasoro algebra},
Vertex operators in mathematics and physics \textbf{3}, 
Springer, (1985) 451--473

\bibitem{kedem.klassen.mccoy.melzer.01}
R Kedem, T R Klassen, B M McCoy, and E Melzer, 
\textit{Fermionic quasi-particle representations for characters of 
$G^{(1)}_1 \times G^{(1)}_1 / G^{(1)}_2$},
Physics Letters \textbf{B 304.3--4} (1993) 263--270,
\texttt{hep-th/9211102}

\bibitem{kedem.klassen.mccoy.melzer.02}
R Kedem, T R Klassen, B M McCoy, and E Melzer, 
\textit{Fermionic sum representations for conformal field theory characters},
Physics Letters \textbf{B 307.1--2} (1993) 68--76,
\texttt{hep-th/9301046}

\bibitem{berkovich}
A Berkovich, 
\textit{Fermionic counting of RSOS states and Virasoro character formulas for 
the unitary minimal series $M \ll v, v+ 1 \rr$: Exact results},
Nuclear Physics \textbf{B 431.1--2} (1994) 315--348,
\texttt{hep-th/9403073}

\bibitem{berkovich.mccoy}
A Berkovich and B M McCoy, 
\textit{Continued Fractions and Fermionic Representations for Characters of 
$M \ll p, p^{\, \prime} \rr$  minimal models}, 
Letters in Mathematical Physics \textbf{37} (1996) 49--66, 
\texttt{arXiv:hep-th/9412030} 

\bibitem{warnaar.01}
S O Warnaar, 
\textit{Fermionic solution of the Andrews-Baxter-Forrester model. 
I. Unification of TBA and CTM methods},
Journal of statistical physics \textbf{82.3-4} (1996) 657-685,
\texttt{hep-th/9501134}

\bibitem{warnaar.02}
S O Warnaar, 
\textit{Fermionic solution of the Andrews-Baxter-Forrester model. 
II. Proof of Melzer's polynomial identities},
Journal of statistical physics \textbf{84.1} (1996) 49--83,
\texttt{hep-th/9508079}

\bibitem{foda.lee.welsh}
O Foda, K S M Lee, and T A Welsh, 
\textit{A Burge tree of Virasoro-type polynomial identities}, 
International Journal of Modern Physics \textbf{A13} (1998) 4967--5012, 
\texttt{arXiv:q-alg/9710025}

\bibitem{foda.lee.pugai.welsh}
O Foda, K S M Lee, Y Pugai, and T A Welsh,
\textit{Path generating transforms}, 
in 
\textit{$q$-Series from a contemporary perspective}, (South Hadley, MA, 1998), 157--186, 
Contemporary Mathematics \textbf{254}, American Mathematical Society, Providence, RI, 2000,
\texttt{arXiv:math/9810043} 

\bibitem{foda.welsh}
O Foda and T A Welsh
\textit{On the combinatorics of Forrester-Baxter models}, 
in
\textit{Physical combinatorics} (Kyoto, 1999), 49--103, Progress in Mathematics \textbf{191}, 
Birkhauser, Boston, MA, 2000,
\texttt{arXiv:math/0002100}

\bibitem{welsh}
T A Welsh,
\textit{Fermionic expressions for minimal model Virasoro characters}, 
Memoirs of the American Mathematical Society (2005) \textbf{175 N 827} 1--160, 
\texttt{arXiv:math/0212154} 

\bibitem{date.jimbo.miwa.okado.01}
A Date, M Jimbo, T Miwa, and M Okado,
\textit{Automorphic properties for local height probabilities for integrable solid-on-sold models},
Physical Review \textbf{ B 35} (1987) 2105--2107

\bibitem{mumford.01}
D Mumford, 
\textit{Tata Lectures on Theta I}, 
Progress in Mathematics \textbf{28} 
(1983) Birkhauser
\texttt{ISBN-10:3764331097},
\texttt{ISBN-13:978-3764331092}

\bibitem{kac.peterson}
V G Kac and D H Peterson,
\textit{Infinite-dimensional Lie algebras, theta functions and modular forms},
Advances in Mathematics \textbf{53.2} (1984) 125--264

\bibitem{date.jimbo.miwa.okado.02}
A Date, M Jimbo, T Miwa, and M Okado,
\textit{Solvable lattice models}, 
in {\em Proceedings in Symposia in Pure Mathematics} \textbf{ 49},
Part I (1989) 295--332

\bibitem{goddard.kent.olive.01}
P Goddard, A Kent, and D Olive,
\textit{Virasoro algebras and coset space models}, 
Physics Letters \textbf{152.1--2 B} (1985) 88--92

\bibitem{goddard.kent.olive.02}
P Goddard, A Kent, and D Olive,
\textit{Unitary representations of the Virasoro and super-Virasoro algebras}, 
Communications in Mathematical Physics \textbf{103.1} (1986) 105--119

\bibitem{date.jimbo.kuniba.miwa.okado.03}
A Date, M Jimbo, A Kuniba, T Miwa, and M Okado,
\textit{Exactly solvable SOS models: Local height probabilities and theta function identities},
Nuclear Physics \textbf{ B 290} [FS20] (1987) 231--273 

\bibitem{date.jimbo.kuniba.miwa.okado.04}
A Date, M Jimbo, A Kuniba, T Miwa, and M Okado,
\textit{Exactly solvable SOS models II: Proof of the star triangle relations 
and combinatorial identities},
in \textit{Advanced Studies in Pure Mathematics} {\bf 16} (1988) 
\textit{Conformal field theory and solvable lattice models}, 17--122 

\bibitem{huse}
D A Huse, 
\textit{Exact exponents for infinitely many new multicritical points},
Physical Review \textbf{B 30} (1984) 3908--3915

\bibitem{gehlen.rittenberg}
G von Gehlen and V Rittenberg,
\textit{The spectra of quantum chains with free boundary conditions and Virasoro algebras},
Journal of Physics \textbf{A}: Mathematical and General \textbf{19.10} (1986) L631--L635

\bibitem{boccara.sarma}
N Boccara and G Sarma,
\textit{Does magnetic surface order exist?},
Journal de Physique Lettres \textbf{35.7-8} (1974) 95--97

\bibitem{burkhardt.guim}
T Burkhardt and I Guim,
\textit{Finite-size scaling of the quantum Ising chain with periodic, free, and antiperiodic boundary conditions},
Journal of Physics \textbf{A}: Mathematical and General \textbf{18.1} (1985) L33--L37

\bibitem{cardy.01}
J L Cardy, 
\textit{Operator content of two-dimensional conformally invariant theories},
Nuclear Physics \textbf{B 270 [FS10]} (1986) 186--204

\bibitem{superstrings.book}
\textit{\lq Superstrings. The first 15 years of superstring theory \rq}, 
Volumes 1 and 2,
J H Schwarz (Editor) World Scientific, Singapore, 1985, 
\texttt{ISBN-10: 9971978679}

\bibitem{cardy.02}
J L Cardy, 
\textit{Effect of boundary conditions on the operator content of two-dimensional 
conformally invariant theories},
Nuclear Physics \textbf{B 275 [FS17]} (1986) 200--218

\bibitem{saleur}
H Saleur,
\textit{Conformal invariance for polymers and percolation},
Journal of Physics \textbf{A}: Mathematical and General \textbf{20.2} (1987) 455--470

\bibitem{saleur.bauer}
H Saleur and M Bauer,
\textit{On some relations between local height probabilities and conformal invariance}, 
Nuclear Physics \textbf{B 320} (1989) 591--624 

\bibitem{hamer.quispel.batchelor}
C J Hamer, G R W Quispel, and M T Batchelor,
\textit{Conformal anomaly and surface energy for Potts and Ashkin-Teller quantum chains},
Journal of Physics \textbf{A}: Mathematical and General \textbf{20.16} (1987) 5677--5693

\bibitem{alcaraz.barber.batchelor}
F C Alcaraz, M N Barber, and M T Batchelor,
\textit{Conformal invariance and the spectrum of the XXZ chain},
Physical Review Letters \textbf{58.8} (1987) 771--774

\bibitem{seaton.nienhuis}
K A Seaton and B Nienhuis,
\textit{Surface critical phenomena and cylindrical partition functions for the CSOS model},
Nuclear Physics \textbf{B 384} (1992) 507--522

\bibitem{foda}
O Foda, 
\textit{Off-ctitical local height probabilities and critical partition functions on a cylinder},
University of Nijmegen preprint (1991) 

\bibitem{shiraishi.kubo.awata.odake}
J Shiraishi, H Kubo, H Awata, and S Odake,  
\textit{A Quantum Deformation of the Virasoro Algebra and the Macdonald Symmetric Functions},
Letters in Mathematical Physics \textbf{38} (1996) 33--51, 
\texttt{arXiv:q-alg/9507034} 

\end{thebibliography}
\end{document}